\documentclass{aa}  

\usepackage{graphicx}
\usepackage{siunitx}
\usepackage{booktabs}
\usepackage{bm}
\usepackage{txfonts}
\usepackage{hyperref}
\hypersetup{colorlinks=true, linkcolor=red, urlcolor=black, citecolor=blue}
\usepackage{array,multirow}
\usepackage{float}
\newcommand*\diff{\mathop{}\!\mathrm{d}}
\usepackage{gensymb}

\usepackage[table, dvipsnames]{xcolor}
\definecolor{bkg}{RGB}{218,229,215}
\usepackage{siunitx}

\newcommand{\oiidoublam}{[\textrm{O}\textsc{ii}]\ensuremath{\lambda3727,3729}}
\newcommand{\oiiiv}{[\textrm{O}\textsc{iii}]\ensuremath{\lambda5007}}
\newcommand{\mgii}{\textrm{Mg}\textsc{ii}~\ensuremath{\lambda2800}}
\newcommand{\feii}{\textrm{Fe}\textsc{ii}}
\newcommand{\feiilist}{\textrm{Fe}\textsc{ii}~\ensuremath{\lambda2344,2374,2382,2586,2600}}
\newcommand{\alii}{\textrm{Al}\textsc{ii}~\ensuremath{\lambda1671}}
\newcommand{\aliiilam}{\textrm{Al}\textsc{iii}~\ensuremath{\lambda1854}, \textrm{Al}\textsc{iii}~\ensuremath{\lambda1862}}
\newcommand{\heiabs}{\textrm{He}\textsc{i}~\ensuremath{\lambda3188}}
\newcommand{\cii}{[\textrm{C}\textsc{ii}]\ensuremath{\lambda2326}}
\newcommand{\ciiidoub}{\textrm{C}\textsc{iii}]\ensuremath{\lambda\lambda1907,1909}}
\newcommand{\oiiDQSETT}{[\textrm{O}\textsc{ii}]\ensuremath{\lambda2470}}
\newcommand{\lya}{\ifmmode {\rm Ly}\alpha \else Ly$\alpha$\fi}
\newcommand{\piemd}{{\tt PIEMD}}
\newcommand{\piemdbgg}{{\tt PIEMD+BGG}}
\newcommand{\spemd}{{\tt SPEMD}}
\newcommand{\spemdbgg}{{\tt SPEMD+BGG}}
\newcommand{\piemdOm}{{\tt PIEMD\_$\Omega_{\rm m}$}}
\newcommand{\spemdOm}{{\tt SPEMD\_$\Omega_{\rm m}$}}
\newcommand{\piemdOmw}{{\tt PIEMD\_$\Omega_{\rm m}w$}}
\newcommand{\spemdOmw}{{\tt SPEMD\_$\Omega_{\rm m}w$}}
\usepackage{natbib}
\bibpunct{(}{)}{;}{a}{}{,} % to follow the A&A style
\DeclareSIUnit\angstrom{\text {Å}}

\defcitealias{Bolamperti2023}{B23}

\begin{document}

   \title{Cosmography from accurate mass modeling of the lens group SDSS J0100+1818: five sources at three different redshifts}

\author{A.~Bolamperti\inst{1,2,3} \thanks{\email{andrea.bolamperti@phd.unipd.it}}
    \and C.~Grillo\inst{4,5}
    \and G.~B.~Caminha\inst{6,7}
    \and G.~Granata\inst{4,8}
    \and S.~H.~Suyu\inst{6,7,9}
    \and \\ R.~Ca\~nameras\inst{7,6,10} 
    \and L.~Christensen\inst{11}
    \and J.~Vernet\inst{3}
    \and A.~Zanella\inst{2}
    %Giovanni, Gabriel, Raoul, Sherry, Lise, Joël, Anita, Aleksi
    }

\institute{Dipartimento di Fisica e Astronomia, Università degli Studi di Padova, Vicolo dell'Osservatorio 3, I-35122 Padova, Italy 
     \and INAF -- Osservatorio Astronomico di Padova, Vicolo dell'Osservatorio 5, I-35122, Padova, Italy \label{inafpd}
     \and European Southern Observatory, Karl-Schwarzschild-Strasse 2, D-85748 Garching bei M\"unchen, Germany
     \and Dipartimento di Fisica, Università degli Studi di Milano, via Celoria 16, I-20133 Milano, Italy
     \and INAF-IASF Milano, via A. Corti 12, I-20133 Milano, Italy
     \and Technical University of Munich, TUM School of Natural Sciences, Physics Department, James-Franck-Stra{\ss}e 1, 85748 Garching, Germany \label{tum}
     \and Max-Planck-Institut f\"ur Astrophysik, Karl-Schwarzschild-Str. 1, D-85748 Garching, Germany \label{mpa}
     \and Dipartimento di Fisica e Scienze della Terra, Università degli studi di Ferrara, via Saragat 1, I-44122 Ferrara, Italy \label{unife} 
     \and Institute of Astronomy and Astrophysics, Academia Sinica, 11F of ASMAB, No.1, Section 4, Roosevelt Road, Taipei 106216, Taiwan 
     \and Aix Marseille Univ, CNRS, CNES, LAM, Marseille, France
     \and Cosmic Dawn Center, Niels Bohr Institute, University of Copenhagen, Jagtvej 128, 2200 N, Copenhagen, Denmark     
}

   \date{Received --; accepted --}

% \abstract{}{}{}{}{} 
% 5 {} token are mandatory
 
  \abstract
  {Systems where multiple sources at different redshifts are strongly lensed by the same deflector allow one to directly investigate the evolution of the angular diameter distances as a function of redshift, and thus to learn about the geometry of the Universe. We present measurements of the values of the total matter density, $\Omega_{\rm m}$, and of the dark energy equation of state parameter, $w$, through a detailed strong lensing analysis of SDSS~J0100+1818, a group-scale system at $z=0.581$ with five lensed sources, from $z=1.698$ to $4.95$. We take advantage of new spectroscopic data from the Multi Unit Spectroscopic Explorer (MUSE) on the Very Large Telescope to securely measure the redshift of 65 sources, including the five multiply imaged background sources (lensed into a total of 18 multiple images) and 19 galaxies on the deflector plane, all employed to build robust strong lensing models with the software {\tt GLEE}. 
  The total mass distribution of the deflector is described in a relatively simple way, and includes an extended halo, the brightest group galaxy (BGG) with a measured stellar velocity dispersion of ($380.5 \pm 4.4) \, \si{km.s^{-1}}$, and fainter members. 
  We measure $\Omega_{\rm m} = 0.14^{+0.16}_{-0.09}$ in a flat $\Lambda$ cold dark matter (CDM) model, and $\Omega_{\rm m} = 0.19^{+0.17}_{-0.10}$ and $w = -1.27_{-0.48}^{+0.43}$ in a flat $w$CDM model. 
  Given the presence of different sources angularly close in projection, we quantify through a multi-plane approach their impact on the inferred values of the cosmological parameters. We obtain consistent median values, with uncertainties for only $\Omega_{\rm m}$ increasing by approximately a factor of 1.5.
  Thanks to the remarkably wide radial interval where the multiple images are observed, ranging from 15 to 77~kpc from the BGG, we accurately measure the total mass profile and infer the stellar over total mass profile of the deflector. They result in a total mass of $(1.55 \pm 0.01) \times 10^{13}$~M$_\odot$ within 50~kpc and a stellar over total mass profile decreasing from $45.6^{+8.7}_{-8.3} \%$ at the BGG effective radius to $(6.6\pm 1.1) \%$ at $R \approx 77$~kpc.   
  Our results confirm that SDSS~J0100+1818 is one of the most massive (lens) galaxies known at intermediate redshift and one of the most distant candidate fossil systems. We also show that group-scale systems that act as lenses for $\geq 3$ background sources at different redshifts enable to estimate the values of the cosmological parameters $\Omega_{\rm m}$ and $w$ with an accuracy that is competitive with that obtained from lens galaxy clusters.

  %In this paper, we enhance the study previously published in \citet{Bolamperti2023} thanks to new data from 5 hours of VLT/MUSE observations (program 0110.A-0248, PI:~A.~Bolamperti). New MUSE observations allow us to 1) confirm the redshift of the source C, that was a free parameter in the previous strong lensing model analysis, 2) identify 19 galaxies at the same redshift as the main deflector, confirming that SDSS~J0100+1818 is likely a fossil system, with a central ultra-massive galaxy surrounded by numerous companions, 3) discover two new lensed systems at high-$z$ that will allow us to refine our lensing models and 4) use this system as a unique laboratory to measure the values of the cosmological parameters $\Omega_{\rm m}$ and $\Omega_\Lambda$, that can be measured only in a few known galaxy-scale systems nowadays. Later, with the kinematic data for the lens that we can infer from MUSE, it will be possible to develop a joint strong lensing+dynamics model, that represents the state-of-the-art in this kind of studies. 
  }


   \maketitle
\section{Introduction}
In the currently accepted $\Lambda$ cold dark matter (CDM) scenario, the Universe is almost flat and expanding, and the expansion is accelerating \citep{Riess1998, Perlmutter1999}. The Universe is composed of baryons and dark matter (DM) for $\approx 30\%$ ($\Omega_{\rm m} \approx 0.3$), and the cosmic acceleration is due to the remaining $\approx 70\%$ ($\Omega_\Lambda \approx 0.7$), represented by the so-called dark energy. Our understanding of dark energy is very poor. We believe that it exerts a negative pressure, and it has an equation of state with $w \approx -1$, where $w$ is defined as the ratio between pressure and energy density, $p/\rho c^2$. The $\Lambda$CDM model successfully describes the Universe at large scales ($\gtrsim 1$~Mpc), but presents difficulties in explaining some properties related to the formation of structures at smaller scales, like the sub-halo structures in galaxy clusters \citep{Grillo2015, Meneghetti2020} and the value of the inner slope of DM halos \citep[e.g.,][]{Gnedin2004, Newman2013b, Newman2013a, Martizzi2012}. Moreover, in extended cosmological models, currently available data cannot accurately measure the values of $\Omega_{\rm m}$ and $\Omega_\Lambda$, and different models can be reconciled with the observations \citep[e.g.,][]{Motta2021}. This motivates the continuous investigation and testing of the $\Lambda$CDM model, through new projects and observations. In this context, the use of different and independent cosmological probes is crucial, as they are subject to different systematics and degeneracies. Thus, they can offer valuable help in investigating the current tensions in cosmology \citep{Verde2019, Moresco2022}. 

Strong gravitational lensing is an extremely powerful tool for extragalactic and cosmological studies \citep[e.g.,][]{Bartelmann2010, Treu2010}. Among the many applications, such as characterizing the total and DM mass distributions of clusters of galaxies \citep{Acebron2022, Bergamini2023b, Granata2023} and galaxies \citep{Vegetti2012, Schuldt2019, Ballard2024} acting as lenses, gravitational lensing can be used to probe the geometry of the Universe. Beside being particularly effective in measuring the value of the current expansion rate of the Universe (the Hubble constant, $H_0$) through the observations of multiply lensed variable sources, such as quasars or supernovae, both on galaxy and cluster scales \citep[e.g.,][]{Refsdal1964, Suyu2017, Grillo2018, Birrer2019, Grillo2020, Wong2020, Rusu2020, Shajib2023, Grillo2024}, it also allows one to measure the values of $\Omega_{\rm m}$, $w$, and $\Omega_{\rm k}$ (the latter parametrizing the curvature of the Universe, $\Omega_{\rm k} = 0 $ in a flat geometry), when kinematic data for lens galaxies are available \citep[e.g.,][]{Grillo2008, Cao2012} or in systems where two or more sources are multiply imaged by the same deflector \citep[][and with clusters of galaxies]{Tu2009, Collett2014, Tanaka2016, Smith2021}. 

Massive clusters of galaxies can produce tens to hundreds of multiple images from several background sources and in the last decade they have been employed to measure the values of $\Omega_{\rm m}$, $\Omega_\Lambda$, $w$, and $\Omega_{\rm k}$ \citep[e.g.,][]{Jullo2010, Caminha2016, Caminha2022, Grillo2024}, also thanks to the advent of very deep Integral Field Spectroscopic (IFS) observations, that represent the most effective way to spectroscopically confirm and discover lensed sources. For instance, the number of spectroscopically confirmed multiple images lensed by the Hubble Frontier Field galaxy cluster MACS~J0416.1-2403 has increased, in less than 10 years, from 10 sources observed in 30 multiple images \citep{Grillo2015} to 88 sources lensed into 237 multiple images \citep{Bergamini2023}. 

Unfortunately, the total mass distribution of clusters of galaxies is usually very complex, and this requires several mass components to properly model it, whose parameters may be degenerate with the cosmological ones. On the other hand, galaxy-scale systems are in general easier to model and the lens can often be described with an effective single total mass profile, but the limited number of background sources makes them prone to be affected by the mass-sheet degeneracy \citep{Schneider2014}. Galaxy or group-scale systems with a larger number ($\geq 3$) of lensed background sources might represent the best compromise between these two regimes to learn about the geometry of the Universe. They allow one to measure the values of $\Omega_{\rm m}$ and $w$, independently of that of $H_0$, but only a few of these systems are known to date and are suitable for cosmological studies \citep[e.g.,][]{Smith2021}. Their analysis, even individually, can offer competitive estimates of the cosmological parameter values, and will pave the way to the exploitation of a larger number of systems of this kind, that are foreseen to be discovered with Euclid and the Vera C. Rubin Observatory - Legacy Survey of Space and Time (LSST) \citep{Collett2014, Li2024}.

\begin{figure}[!t]
   \centering
   \includegraphics[width=8cm]{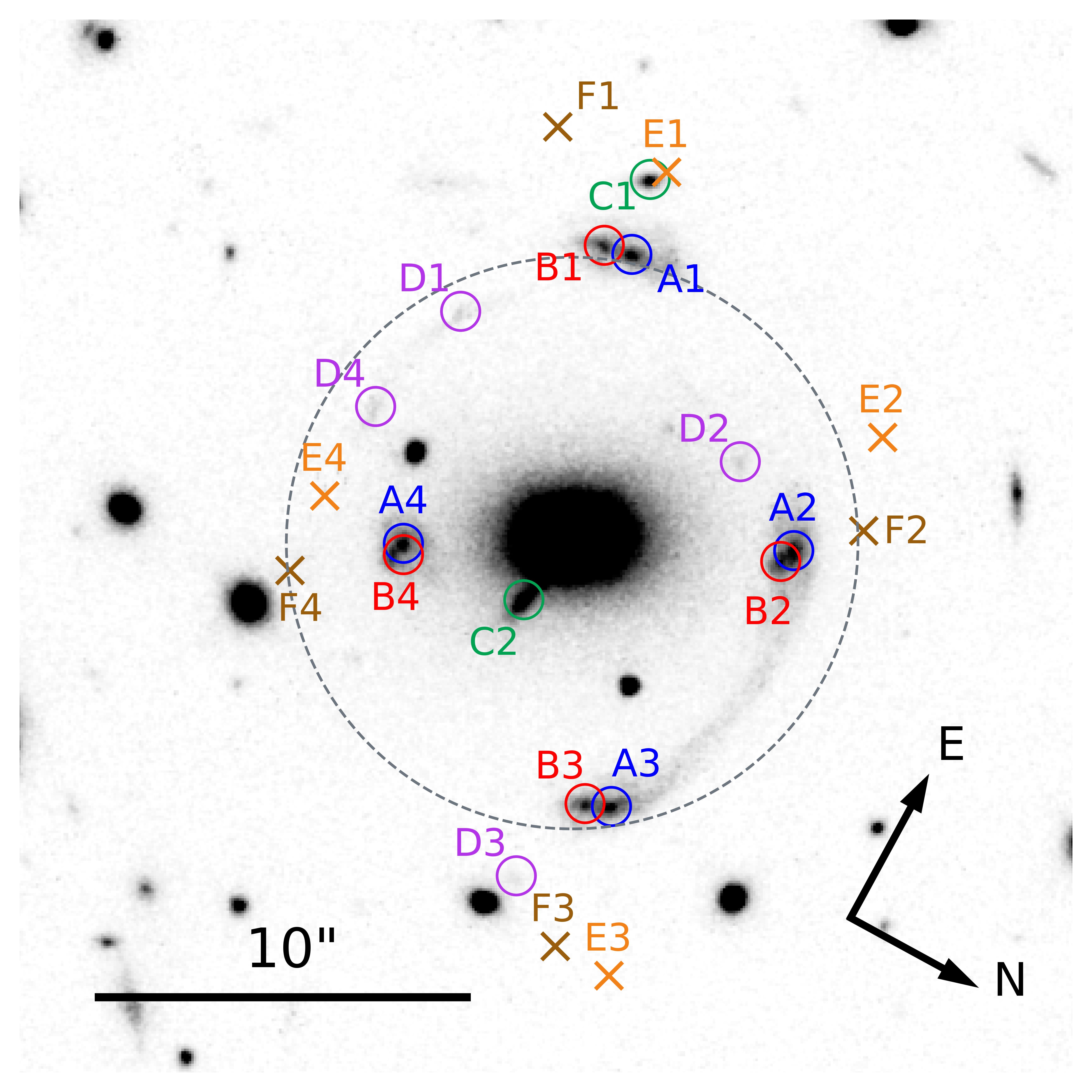}
      \caption{HST F160W image of the SDSS~J0100+1818 strong lensing system studied in this work. Letters label multiple images of the same background source. E and F do not show continuum in the HST image and the orange and brown crosses mark the position of their \lya\ emission line peaks, as detected with MUSE. The grey dashed circle represents an aperture of 50~kpc, the approximate average Einstein radius of the system.}
         \label{fig:labels}
   \end{figure}

In this paper, we extend the study published in \citet{Bolamperti2023} (hereafter, \citetalias{Bolamperti2023}) on SDSS~J010049.18+181827.7 (hereafter, SDSS~J0100+1818), a strong lensing system (Fig.~\ref{fig:labels}) included in the Cambridge And Sloan Survey Of Wide ARcs in the skY (CASSOWARY) survey \citep{Belokurov2009, Stark2013} as a candidate fossil system at $z=0.581$ \citep{Johnson2018}. In our previous work, we developed a strong lensing model of the system from the observed positions of the four multiple images of sources A and B (spectroscopically confirmed) and of the two multiple images of source C (with its redshift as a free parameter), and from the extended surface brightness distributions of the multiple images from the Hubble Space Telescope (HST) data. We employed the best-fit models to measure the cumulative total mass profile of the deflector, disentangle the DM and baryonic mass distributions, and to reconstruct the background sources. Now, we leverage on new data taken with the Multi Unit Spectroscopic Explorer (MUSE) on the Very Large telescope (VLT). These IFS observations allow us to measure the redshift of source C, discover two additional strongly lensed objects (E and F), spectroscopically confirm the group members in the MUSE field of view, and measure the stellar velocity dispersion profile of the brightest galaxy. We develop enhanced strong lensing models by including this information, that also allow us to measure the posterior probability distributions of the cosmological parameters $\Omega_{\rm m}$ and $w$. 

This paper is organized as follows. In Section~\ref{sec:data}, we summarize the currently available data for SDSS~J0100+1818, focusing on the MUSE data reduction and spectra extraction. In Section~\ref{sec:system}, we describe the SDSS~J0100+1818 system, characterizing the multiply imaged sources A, B, C, E, and F (excluding D that is not a secure system),  the deflector with its velocity dispersion profile, and the group members. In Section~\ref{sec:lensing_model}, we describe the enhanced strong lensing models developed and the relative results. In Section~\ref{sec:lensing_model_cosmo}, we show the strong lensing models with variable values of the cosmological parameters $\Omega_{\rm m}$ and $w$. We discuss the results in Section~\ref{sec:discussion}, and draw conclusions in Section~\ref{sec:conclusions}. 
Unless differently specified, and in Section~\ref{sec:lensing_model_cosmo}, throughout this work we assume $H_0 = 70 \, \si{km.s^{-1}.Mpc^{-1}}$, $\Omega_{\rm m} = 0.3$ and $\Omega_\Lambda = 0.7$. In this model, $1$ arcsec corresponds to a linear size of $6.585$~kpc at the deflector redshift of $z = 0.581$. All magnitudes are given in the AB system \citep{Oke1974} and are measured in the HST F160W filter, unless differently specified.

\section{Observations and data reduction}
\label{sec:data}
\subsection{Ancillary data}
SDSS~J0100+1818 benefits from a large sample of both photometric and spectroscopic observations, detailed, with the adopted data reduction procedures, in \citetalias{Bolamperti2023}. In particular, we observed SDSS~J0100+1818 with the HST Wide Field Camera 3 (WFC3; program GO-15253; PI:~R.~Ca\~nameras), spending one orbit in each of the two F438W and F160W filters, resulting in PSF FWHMs of 0.086\arcsec\ and 0.187\arcsec\ in F438W and F160W, respectively.
We complemented photometric observations with VLT/X-Shooter \citep{Vernet2011} spectroscopy (program 091.A-0852, PI:~L.~Christensen), obtaining the spectrum of the main deflector and of three out of four multiple images of the families A and B, being able to spectroscopically confirm them and to measure the lens stellar velocity dispersion. 

\subsection{Integral field spectroscopy observations with VLT/MUSE}
We added IFS data of SDSS~J0100+1818, by observing it for 4.8 hours with VLT/MUSE, from October to December 2022 (program 0110.245R, PI:~A.~Bolamperti). The observations were taken with seeing $\lesssim 1$\arcsec, clear sky conditions, airmass $\lesssim 1.4$ and with the Wide Field mode, resulting in a 1\arcmin\,$\times$\,1\arcmin\ field of view and a spatial sampling of 0.2\arcsec\ pix$^{-1}$. We divided the observations into six different observation blocks (OBs) composed of two exposures of 1440$\,$s each. Each exposure was taken with four different rotation angles (90\degree$\,$apart) and applying small offsets to the center, to optimize the final data quality. 

We reduced the data with the standard MUSE pipeline v2.8.9 \citep{Weilbacher2020}, following the procedure detailed in \citet{Caminha2019}, through the ESO Recipe Execution Tool \citep[\texttt{EsoRex};][]{esorex} pipeline. In summary, we corrected all the raw exposures using the associated \texttt{BIAS}, \texttt{FLAT}, and \texttt{ILLUMINATION} calibrations. We then applied wavelength and flux calibrations and created a \texttt{PIXTABLE} relative to each exposure. 
%
%\gbc{Did you use the \texttt{--autocalib=deepfield} for the autocalib option to reduce the background variation in each of the 24 IFUs? If yes you should mention this here.}
%
At the end, we combined all of them into a final stacked datacube, and defined the astrometry with respect to the HST F160W image. To enhance the background sky subtraction, we made use of the Zurich Atmosphere Purge \citep[\texttt{ZAP};][]{Soto2016} tool. 

The resulting datacube spans in wavelength from $4750 \, \si{\angstrom}$ to $9350 \, \si{\angstrom}$, with a constant $1.25 \, \si{\angstrom}$ pix$^{-1}$ sampling. The final reduced MUSE data cube has a median PSF FWHM of 0.8\arcsec, covers a square of approximately 400 kpc on a side at the redshift of the main deflector, and it is centered on it. This pointing includes all the multiple images previously identified, as well as the most likely group members from our photometric analysis. 

\begin{figure}
   \centering
   \includegraphics[width=0.95\columnwidth]{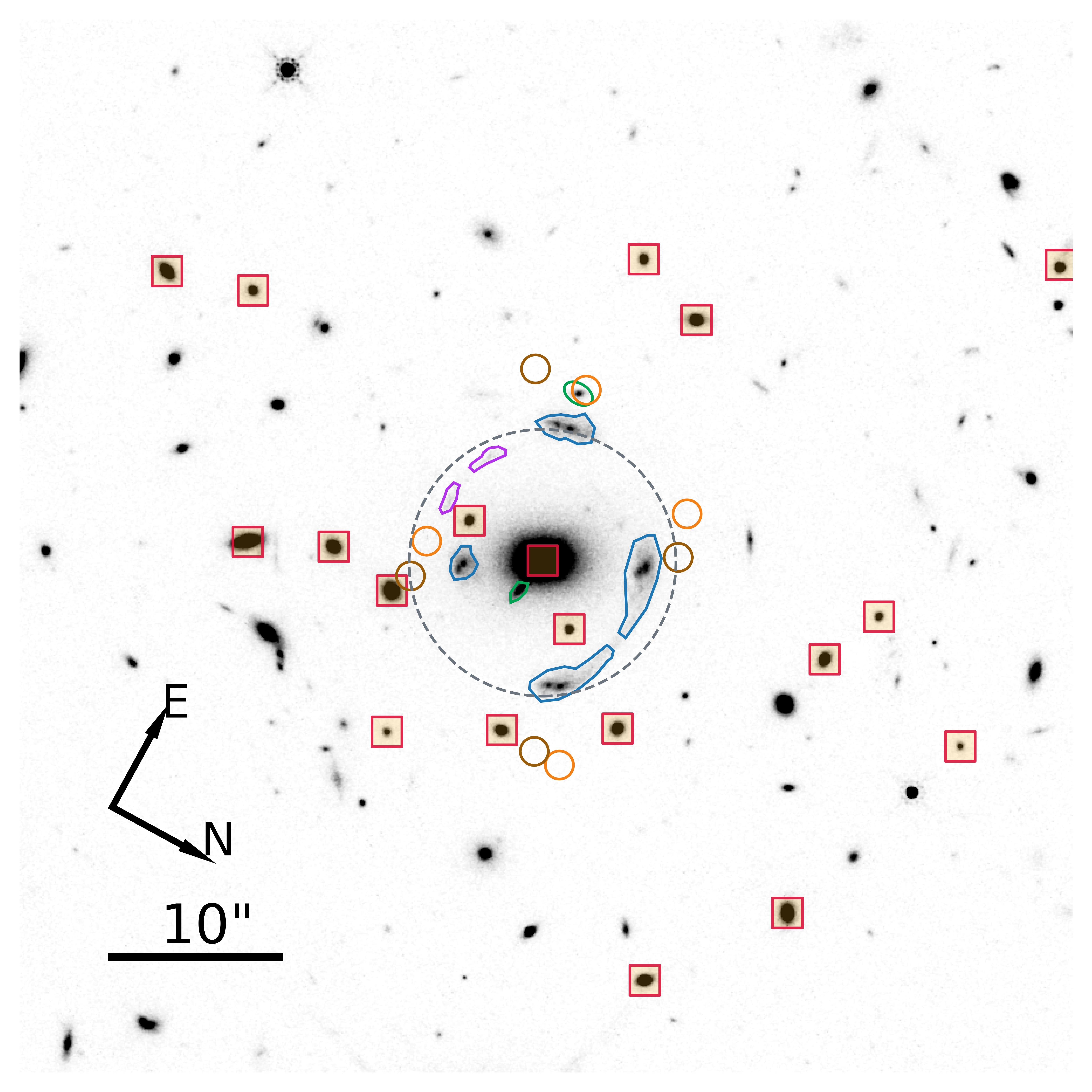}
      \caption{Group members (squares) identified as those galaxies whose redshift is within $0.581 \pm 0.011$, equivalent to the mean redshift of the group with a spread of $2000 \, \si{km.s^{-1}}$ (rest-frame). 
      The colored regions represent the apertures within which we extracted the spectra of the multiple images of A and B (blue), C (green), D (purple), E (orange) and F (brown). The spectra of E and F, that do not present HST-detected continuum, are extracted within circular apertures with radius of 0.8\arcsec. The grey dashed curve shows a circle with a radius of 50~kpc, which represents approximately the average Einstein radius of the system.}
         \label{fig:regions}
   \end{figure}

We identified and measured the source redshifts following the multi-step procedure described in \citet{Caminha2019}. Firstly, we ran {\tt SExtractor} \citep[{\tt v2.28.0},][]{Bertin1996} on a cutout of the HST F160W image that covers the final MUSE datacube field of view (shown in Fig.~\ref{fig:regions}), to detect the position of all the sources. We then extracted a spectrum from the datacube within a circular aperture with radius of 0.8\arcsec\ centered on each detected position. This aperture was chosen to represent a good compromise between collecting as much signal as possible and not including nearby contaminants in the most crowded regions. For some particular systems of distorted arcs hosting multiple images, or with faint and extended objects, we assumed specifically-designed regions to extract the spectra, that allowed us to obtain a better $S/N$. We complemented this catalog by adding sources that present clear emission lines in the MUSE datacube, but are not detected in the HST F160W continuum, through visual inspection and the use of the Cube Analysis and Rendering Tool for Astronomy \citep[\texttt{CARTA};][]{carta_2021} tool. We extracted their spectra within circular apertures with radius of 0.8\arcsec\ centered on the luminosity peak of the detected emission lines.

We measured the redshift values of the objects in the catalog with a dual automatic and visual procedure, identifying clear spectral features like emission and/or absorption lines, and continuum breaks. We made use of the software \texttt{Marz} \citep{Hinton2016Marz}, that performs an automatic fitting of each spectrum with different templates. We found the automatic procedure reliable in the high $S/N$ regime (e.g., to identify the bright group members), while we visually inspected the faint objects or those with only emission lines detected. We assigned for each redshift measurement a quality flag (QF), defined to be QF$\,=1$ for non-reliable measurements, QF$\,=2$ for possible measurements based on faint spectral features, QF$\,=3$ for secure measurements from more than one absorption and/or emission lines, and QF$\,=9$ for measurements based on a single narrow line emission. Most of the objects with QF$\,=9$ present features that allowed us to identify their nature, like the \lya\ line or a doublet, and their reliability is similar to that of the QF$\,=3$ objects \citep[see][]{Grillo2015, Balestra2016}. The catalog used in this work contains 133 redshift measurements with QF$\,\geq 2$ (65 with QF$\,\geq 3$), divided into 3 (3) stars, 22 (19) group members, 27 (10) foreground galaxies, 80 (32) background galaxies and 1 (1) background quasar. 

\section{The SDSS~J0100+1818 system}
\label{sec:system}
SDSS~J0100+1818, (RA, dec$)=($01:00:49.18,~+18:18:27.79), was introduced in a later release of the CASSOWARY catalog. In our previous work \citepalias{Bolamperti2023}, we determined a robust lens redshift, $z=0.581$, and a stellar mass value of $(1.5 \pm 0.3) \times 10^{12}$~M$_{\odot}$ for the most luminous galaxy. We assumed different mass distributions for the main lens and, through a strong lensing analysis, we measured the total mass profile of the deflector. We have recently spotted a flaw in our code that caused the deflector total mass profile to be slightly underestimated. In the following, and in particular in Section~\ref{sec:comparison_with_2023}, we refer to the corrected values. They consist in a total mass value of ($1.16\pm 0.01) \times 10^{13}$~M$_{\odot}$ within 42~kpc and a stellar over total mass fraction of ($38 \pm 9$)\% at the half-light radius ($R_{\rm e} = 9.3$~kpc) of the main lens galaxy. These values are consistent with the previous estimates, and do not affect the results and discussion presented in \citetalias{Bolamperti2023}, that remain valid.
%, and was not initially considered in the spectroscopic confirmation program conducted by \citet{Stark2013}. Consequently, we followed it up with X-Shooter observations to accurately determine redshifts for both the primary lens elliptical galaxy and the lensed sources. Previously \citep{Bolamperti2023}, we determined the most robust lens redshift, $z=0.581$, based on the prominent absorption lines found in the rest-frame optical spectrum. We measured a stellar velocity dispersion value of $\sigma_\star=(451 \pm 37)$  km~s$^{-1}$ and a best-fit stellar mass value of $(1.5 \pm 0.3) \times 10^{12}$~M$_{\odot}$. 

%

\subsection{Multiple images of A and B}
\begin{figure*}
   \centering
   \includegraphics[width=0.95\textwidth]{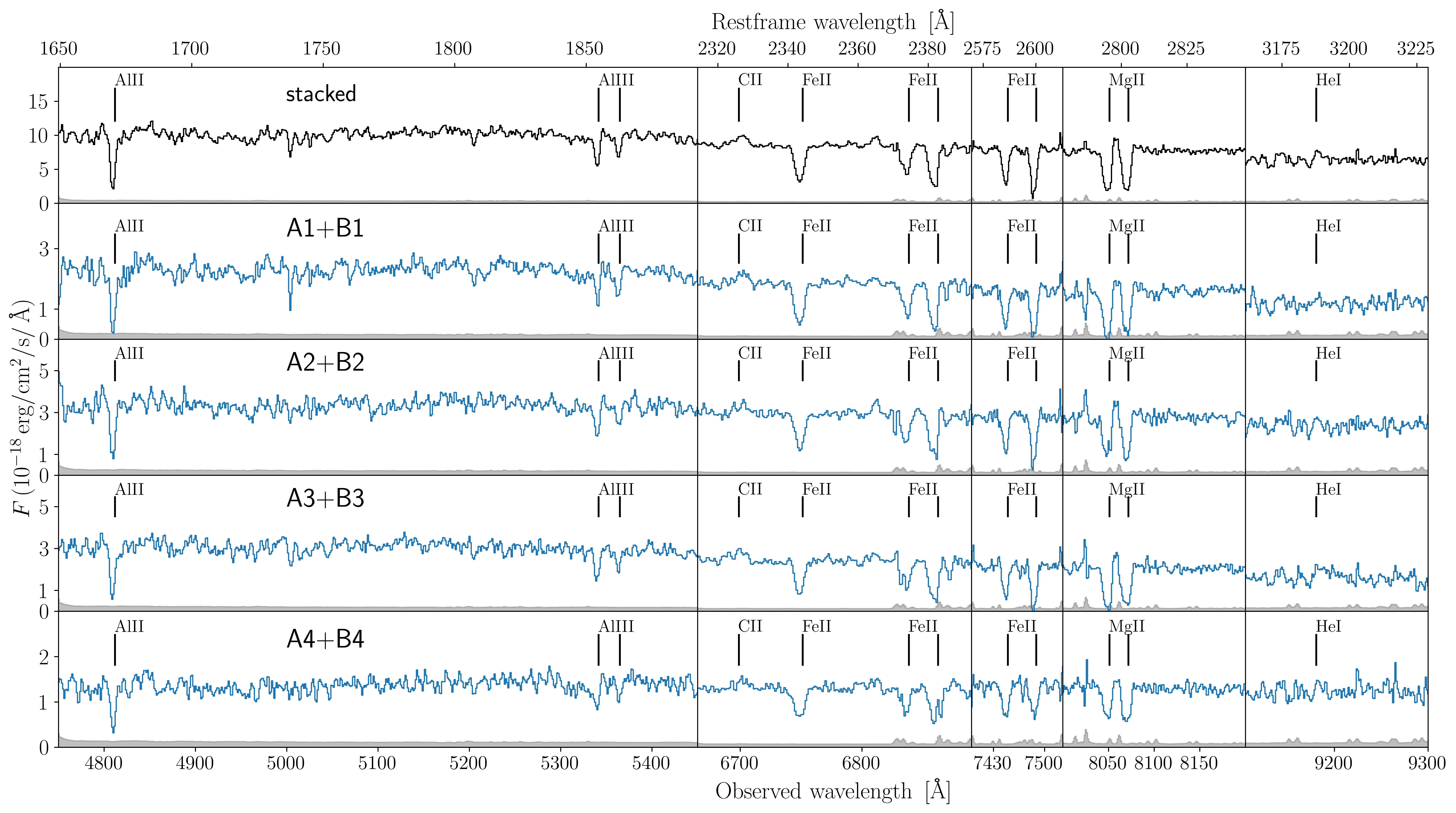}
      \caption{MUSE 1D spectra of the multiple images A1+B1, A2+B2, A3+B3, A4+B4, and stacked (top row), extracted within the blue apertures highlighted in Fig.~\ref{fig:regions}. The spectra are smoothed with a $\sim 2.5 \, \AA$ boxcar filter. We zoom in around five wavelength ranges where we observe the main absorption and emission lines used to confirm the redshift measurement of 1.880. The blue lines represent the observed spectra in units of $10^{-18}$ erg s$^{-1}$ cm$^{-2}$ $\AA^{-1}$, and the gray regions indicate the 1$\sigma$ uncertainties.}
         \label{fig:specAB}
   \end{figure*}

We confirm the joint redshift measurement of $z=1.880$ for the two source components forming image families A and B (Fig.~\ref{fig:labels}) that we found in \citetalias{Bolamperti2023}. Previously, based on X-Shooter data, we performed a joint analysis of the targeted multiple images, A1/B1, A3/B3, and A4/B4. We made use of lines detected at about $10740 \, \AA$ in the binned 2D spectra, whose width is consistent with the \oiidoublam\ doublet, of a faint detection of \oiiiv\ in A3/B3, and of the lack of additional line detections over the spectral range covered by X-Shooter. 
In the MUSE cube, we extract the spectra over four regions, shown in Fig.~\ref{fig:regions}, that include the compact and extended emission from all four images. We detect the \mgii\ doublet at about $8065 \, \AA$, five absorption features of \feii\ (\feiilist), three of Al (\alii, \aliiilam), and a tentative detection of \heiabs\ and \cii.  
These features are shown, for each of the A1/B1, A2/B2, A3/B3, A4/B4 images and stacked, in Fig.~\ref{fig:specAB}. 

\subsection{Multiple images of C}
\begin{figure*}
   \centering
   \includegraphics[width=0.95\textwidth]{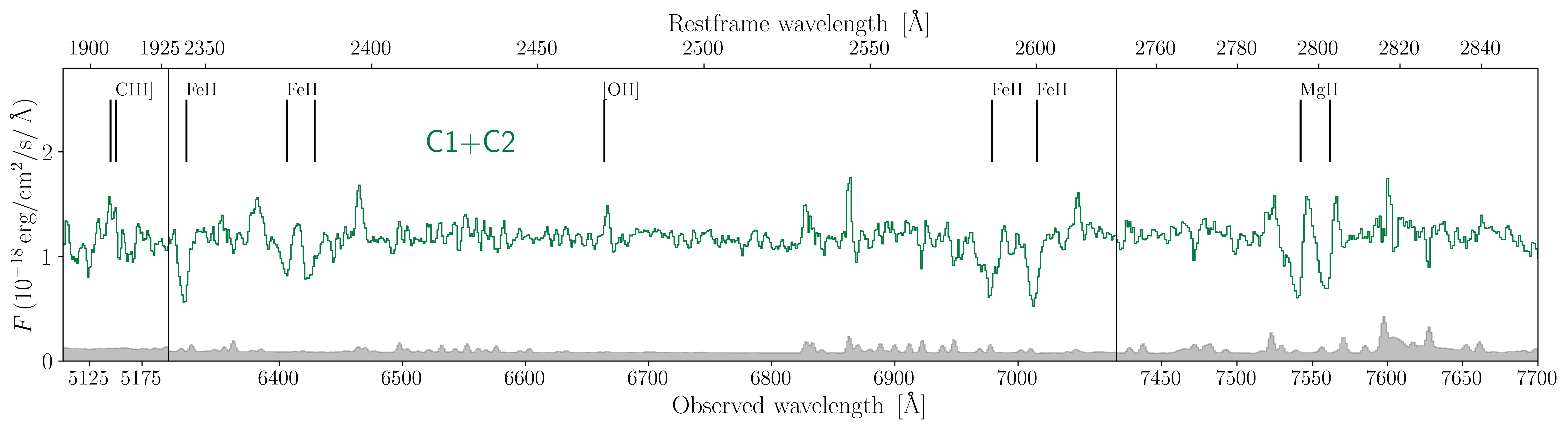}
      \caption{MUSE 1D spectrum of the multiple images C1 and C2 (stacked), extracted within the green apertures highlighted in Fig.~\ref{fig:regions}. The spectrum is smoothed with a $\sim 2.5 \, \AA$ boxcar filter. We zoom in around three wavelength ranges where we observe the main absorption and emission lines used to confirm the redshift measurement of 1.698. The green line represents the observed spectra in units of $10^{-18}$ erg s$^{-1}$ cm$^{-2}$ $\AA^{-1}$, and the gray regions indicate the 1$\sigma$ uncertainties.}
         \label{fig:specC}
   \end{figure*}

By combining observations in the HST F160W and F438W filters with the first strong lensing models of the system, we previously identified another candidate lensed background source, with two multiple images labeled as C1 and C2 in Fig.~\ref{fig:labels}. Since C1 and C2 have similar colors and their positions were correctly predicted by our initial strong lensing models, it was highly likely that a single source was observed multiple times. Although we lacked spectroscopic confirmation, as they were not included in the X-Shooter pointing, we considered the redshift value of source C as a free parameter in the strong lensing analysis. The two models we assumed as the most reliable ones in \citetalias{Bolamperti2023} predicted, in the point-like approximation and in the extended source reconstruction, best-fit redshift values of respectively 1.72 and 1.69. They also showed that the value of $z_{\rm C}$ was degenerate with those of other parameters, mainly the Einstein radius ($\theta_\mathrm{E}$) and the 3D logarithmic total mass density slope ($\gamma^\prime$) of the deflector.
We extract the spectra of C1 and C2 from the apertures shown in Fig.~\ref{fig:regions}, and measure a redshift value of $1.698$, from the detection of the \mgii\ at about $7550~\AA$, with some evidence of a P-Cygni line profile of the doublet, five absorption features of \feii\ (\feiilist), and the emission lines of \ciiidoub\ and \oiiDQSETT, shown in Fig.~\ref{fig:specC}.

\subsection{Multiple images of E and F}
\begin{figure*}
   \centering
   \includegraphics[width=0.95\textwidth]{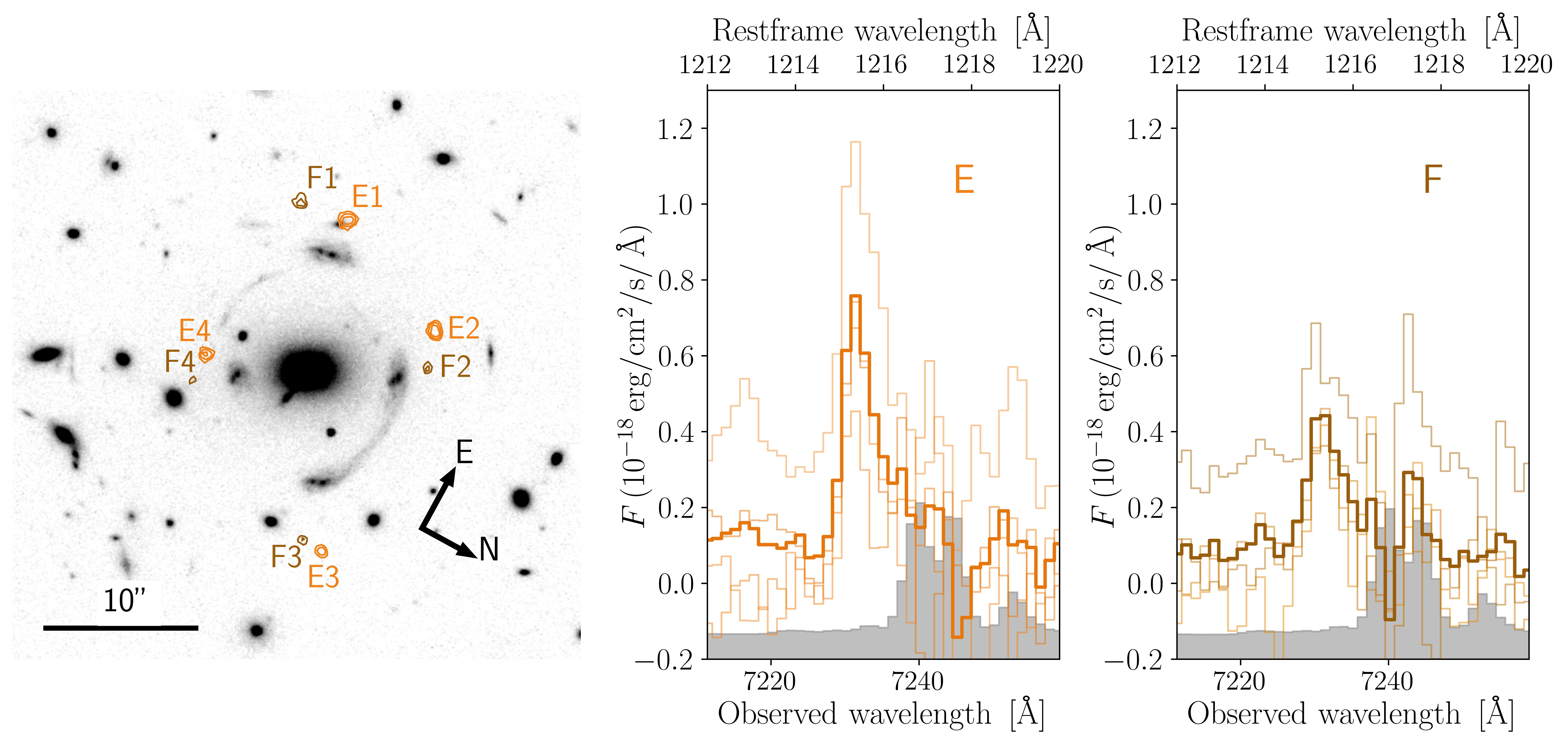}
      \caption{Left: contours from a continuum-subtracted narrow band image, centered on the emission line detected at $\approx 7231 \, \AA$, identified as the \lya\ emission at $z=4.95$. The contours are superimposed on the HST F160W image, to show that continuum counterparts are not detected. Center: zoom-in of the MUSE 1D spectra of the multiple images E1, E2, E3, E4, and stacked (thicker line), around the detected emission line. Right: same as in the center for F1, F2, F3 and F4. The spectra are extracted within the orange (E) and brown (F) apertures in Fig.~\ref{fig:regions}, consisting of circular apertures with radius of 0.8\arcsec, and are smoothed with a $\sim 2.5 \, \AA$ boxcar filter.}
         \label{fig:specEF}
   \end{figure*}

With MUSE we identify two additional multiply imaged background sources, labeled as E and F in Fig.~\ref{fig:labels}, that do not show any stellar continuum counterpart in HST F160W. They are detected through a strong emission line at $\approx 7231 \, \AA$, as it often happens in deep MUSE observations of similar \citep[e.g.,~][]{Collett2020} and blank \citep[e.g.,~][]{Bacon2023} fields. Based on the asymmetric line profile with a clear blue cutoff, this line is interpreted as the \lya\ line at $z=4.95$. E and F have four multiple images each, whose spectrum around the observed emission line is shown in Fig.~\ref{fig:specEF}. Their position coordinates are reported in Table~\ref{tab:multiple_images}.

\begin{table}
\caption{Right ascension (RA) and declination (dec), redshift ($z$), and distance in projection from the brightest group galaxy (BGG) center ($d$) of the detected multiple images.}
    \centering
    \begin{tabular}{ccccc}
    \hline
    \hline
        ID & RA & dec & $z$ & $d$ [\arcsec] \\
        \hline
        BGG & 01:00:49.18 & +18:18:27.79  & 0.581 & 0.00 \\
        A1 & 01:00:49.71 & +18:18:25.50  & 1.880 & 7.54 \\
        A2 & 01:00:49.36 & +18:18:33.05 & 1.880 & 5.86 \\
        A3 & 01:00:48.78 & +18:18:32.06 & 1.880 & 7.37\\
        A4 & 01:00:49.03 & +18:18:23.85 & 1.880 & 4.56\\
        B1 & 01:00:49.69 & +18:18:24.74 & 1.880 & 7.67\\
        B2 & 01:00:49.33 & +18:18:32.89 & 1.880 & 5.54\\
        B3 & 01:00:48.76 & +18:18:31.40 & 1.880 & 7.23\\
        B4 & 01:00:48.99 & +18:18:23.75 & 1.880 & 4.86 \\
        C1 & 01:00:49.84 & +18:18:24.98 & 1.698 & 9.60\\
        C2 & 01:00:49.04 & +18:18:27.38 & 1.698 &2.25 \\
        E1 & 01:00:49.87 & +18:18:25.27 & 4.95 & 10.01\\
        E2 & 01:00:49.63 & +18:18:33.69 & 4.95 & 8.64 \\
        E3 & 01:00:48.50 & +18:18:34.15 & 4.95 & 11.70\\
        E4 & 01:00:49.03 & +18:18:21.41 & 4.95 & 6.70\\
        F1 & 01:00:49.85 & +18:18:22.15 & 4.95 & 10.91\\
        F2 & 01:00:49.46 & +18:18:34.43 & 4.95 & 7.72\\
        F3 & 01:00:48.50 & +18:18:32.52 & 4.95 & 10.90\\
        F4 & 01:00:48.88 & +18:18:21.55 & 4.95 & 7.59\\
        \bottomrule
    \end{tabular}
    \tablefoot{The coordinates of the multiple images of families A, B, and C are measured with \texttt{SExtractor} in the HST F160W image, while the positions of the multiple images of the E and F sources, whose continuum is not detected in HST, are taken as the brightest pixels in a narrow-band image centered around their \lya\ emission lines.
    }
    \label{tab:multiple_images}
\end{table}

\subsection{Discussion on system D}
We identify an additional system with four multiple images, labeled as D1-4 in Fig.~\ref{fig:labels}. D1 and D4 are distorted into an extended arc South-East to the main elliptical deflector, D2 is faint in the North-East direction, and D3 lies between B3 and F3 in the West direction, angularly close to a group member with a F160W magnitude of 21.32 mag. We extracted the D spectrum from an elongated region that includes the D1/D4 arc, shown in Fig.~\ref{fig:regions}. We also extracted the spectra of D2 and D3 within circular apertures, but they revealed to be too faint and too contaminated, respectively. We do not detect any clear spectral feature that allows us to obtain a secure redshift measurement for D. The lack of strong emission lines in MUSE and the positions of the multiple images suggest that D lies at a redshift between approximately 1.5 and 2, but further observations will be needed to confirm the nature of this system.
%
%We identify a single emission line at the red end of the MUSE wavelength coverage, that we interpret as the \oiidoublam\ doublet, that is also consistent with a tentative detection of the \mgii\ doublet at approximately 7000 \AA. These features are shown in the appendix, in Fig.~\ref{fig:specD}. \\
%measure $z_D = 1.503$, labeled with QF$=2$, given that we identify a single emission line at the red end of the MUSE wavelength coverage, that we interpret as the \oiidoublam\ doublet. At this redshift, we also tentatively observe the \mgii\ doublet at approximately 7000 \AA. These features are shown in the appendix, in Fig.~\ref{fig:specD}. \\
%
The D system was hypothesized also in our previous study, but did not enter in the analysis. We anticipate here that we continue to exclude the multiple images of source D in the strong lensing modeling described in the following for different reasons. The lack of a secure redshift measurement, the fact that it is not possible to clearly identify a position and a brightest pixel for image D2, and the position of D3 that is strongly perturbed by the mass distribution of a group member angularly very close, make the inclusion of family D strongly uncertain. Furthermore, our simple best-fit strong lensing model, detailed in Section~\ref{sec:lensing_model}, that can reproduce very well the positions of the spectroscopically confirmed multiple images, currently disfavors the inclusion of this system. To improve on that, we would need to include additional mass components, that are not fully justified by the unsecure nature of D. For these reasons, we leave the inclusion of system D for future strong lensing modeling efforts, should its redshift and its spatial positions be confirmed with deeper observations.

%We tried to optimize the values of the parameters of the mass distribution of the deflector with only A, B, C, E, and F, and then adding D, and the goodness of the models strongly decreased (rms, Eq.~\ref{eq:rms}, increasing by a factor of 2). \shs{The previous sentence seems too early and too technical in the text for a reader not familiar with strong lens modeling to understand.  Maybe move this technical part to lens modeling section, and just state here that "Furthermore, our strong lens model currently disfavors the inclusion of system D based on the noisy observables."} 

\subsection{Group members}
In \citetalias{Bolamperti2023}, given the large estimated value of the deflector total mass, we considered the possibility of SDSS~J0100+1818 being a group of galaxies, and we found 53 galaxies in a $2.5\arcmin \times 2.5\arcmin$ field of view with photometric redshifts consistent, at the 2$\sigma$ level, with the spectroscopic redshift of the main elliptical galaxy. 
Thanks to the MUSE data, we securely select and consider as part of the same group all the galaxies with spectroscopic redshift in the $0.581 \pm 0.011$ range, corresponding to the mean redshift of the overdensity at $z \approx 0.6$, with a spread of $\pm 2000 \, \si{km.s^{-1}}$ (rest-frame). With this method, we select 19 group members (with QF$\,\geq 3$), including the main elliptical galaxy, hereafter referred to as brightest group galaxy (BGG), whose positions are shown in Fig.~\ref{fig:regions}. They are distributed through the entire MUSE field of view. The closest and the furthest group members lie approximately 4\arcsec\ (27~kpc) and 33\arcsec\ (220~kpc) away in projection from the BGG, respectively. Three group members are located within 7\arcsec\ (approximately the physical Einstein radius of system AB), eight within 15\arcsec, and thirteen within 20\arcsec. We measure their F160W Kron magnitudes with {\tt SExtractor}, ranging from 17.05 mag to 22.96 mag. We note that the second brightest member has a magnitude of 19.94, almost three magnitudes fainter than the BGG.    

\subsection{Velocity dispersion profile}
\label{sec:kinematics}
\begin{figure}
   \centering
   \includegraphics[width=0.98\columnwidth]{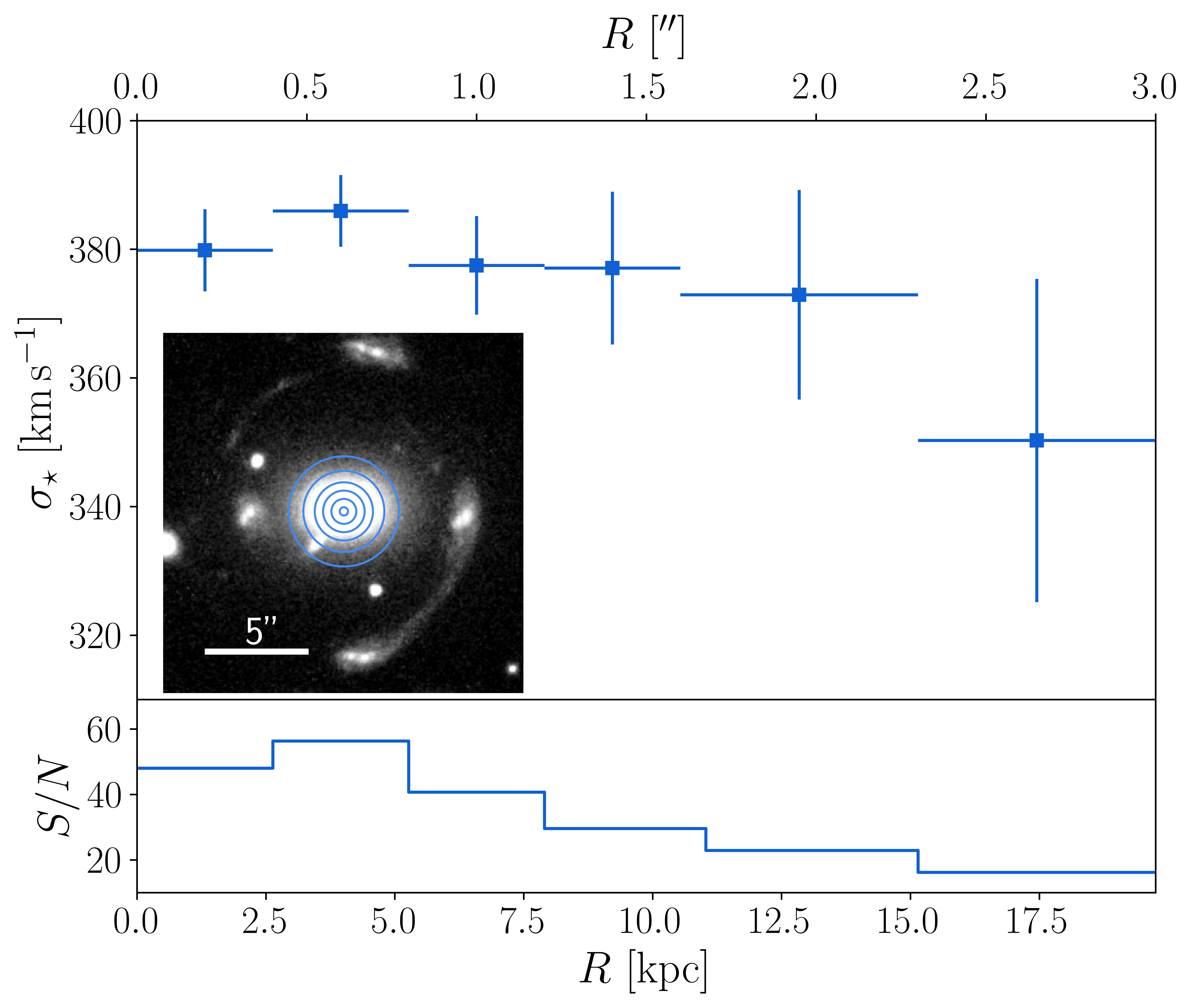}
      \caption{Stellar velocity dispersion profile of the main deflector. The apertures (the first 0.4\arcsec-radius aperture and then different annuli with the same center) used to extract the spectra relative to each bin are shown in the inset. In the bottom panel, we show the spectral $S/N$ of each bin, that is larger than 15 in all of them, peaking at around 60. Given the value of the observational PSF FWHM and of the chosen bin widths, the $\sigma_\star$ values of the profile are partially correlated.}
         \label{fig:sigma_profile}
   \end{figure}

With the new MUSE data, we are able not only to confirm the extremely large stellar velocity dispersion value, $\sigma_\star$, that distinguishes the main deflector to be one of the most massive galaxies known at intermediate redshift, but also to measure a stellar velocity dispersion profile, considering the first aperture of 0.4\arcsec\ of radius and then different annuli with the same center (see Fig.~\ref{fig:sigma_profile}). The spectral $S/N$ is larger than 15 in all the selected bins and the corresponding velocity dispersion values are partially correlated, given the observational PSF FWHM. We measured the value of the stellar line-of-sight velocity dispersion of each bin following the procedure presented in \citet{Granata2023}. We used \texttt{pPXF} \citep[penalized pixel-fitting,][]{Cappellari04,Cappellari17,Cappellari23} to perform a full-spectrum fit, comparing the observed spectra with a combination of stellar templates. These were chosen from a set of 463 UVB stellar spectra with $S/N>100 \, \mathrm{\r{A}^{-1}}$ from the X-Shooter Spectral Library (XSL) DR2 \citep{gonneau20}, de-graded to the instrumental resolution of MUSE, and then combined with additive 12th-degree Fourier polynomials and convolved with a Gaussian line-of-sight velocity distribution. The obtained global stellar velocity dispersion presents a value that is consistent, but systematically lower than the previous estimate of ($451 \pm 37) \, \si{km.s^{-1}}$ with X-Shooter \citepalias{Bolamperti2023}. The difference can be explained mainly by the different $S/N$ of the two datasets (significantly higher for MUSE). %, in addition to the different spectral resolution of the instruments and stellar templates adopted to model the spectra. 
In the following, we will refer to the MUSE estimate, given that it is more robust and will moreover result consistent with the BGG mass obtained from the strong lensing analysis (see Sections~\ref{sec:lensing_model} and \ref{sec:lensing_model_cosmo}).

\section{Lensing modeling}
\label{sec:lensing_model}
We performed our strong lensing modeling with the Gravitational Lens Efficient Explorer \citep[\texttt{GLEE};][]{Suyu2010} software, that allows one both to optimize the parameter values and to infer their probability distributions, through Bayesian analyses like simulated annealing and Markov chain Monte Carlo (MCMC), respectively. \texttt{GLEE} supports several types of mass and light profiles, whose parameters can be optimized by reproducing the observed positions, fluxes, and/or time delays of multiple images. In this study, we employed the positions of the 18 multiple images from the five background sources A, B, C, E, and F, listed in Table~\ref{tab:multiple_images}. For A, B, and C, we considered each multiple image position as that of its brightest pixel in the HST F160W image, with an uncertainty of one HST pixel (0.066\arcsec). For E and F, that do not show continuum in the HST images, we considered the  brightest pixel in a narrow-band image centered around the \lya\ emission line. For these images, we adopted a positional uncertainty of 0.15\arcsec. In this work, we only performed a point-like source modeling, given that the images of the sources E and F do not present any HST continuum. We leave the extended source modeling approach for a future work, where we will treat properly the weighing of a combination of extended and point-like sources.

The best-fit values of the parameters of each model were estimated with a simulated annealing technique, structured in several steps, first minimizing the $\chi^2$ value on the source plane, and then on the deflector plane. We assumed different total mass profiles for the deflector, and evaluated their goodness by considering different statistical estimators, often employed in similar strong lensing studies \citep[see, e.g.,][]{Acebron2017, Mahler2018}. We took into account: (1) the number of degrees of freedom (ndof), defined as the difference between the number of observables and the number of free parameters, (2) the value of the minimum $\chi^2$, defined as 
\begin{equation}
    \label{eq:chi2}
    \chi^2 = \sum_{i=1}^N \frac{\left|\bm{\theta}_i^\mathrm{obs}-\bm{\theta}_i^\mathrm{pred}\right|^2}{\sigma^2_i} \, ,
\end{equation}
where $\bm{\theta}_i^\mathrm{obs}$ are the positions of the observed images, $\bm{\theta}_i^\mathrm{pred}$ are their positions as predicted by the model, $N$ is the total number of multiple images and $\sigma_i$ is the positional uncertainty relative to the $i$-th image, (3) the value of the root-mean-square difference (rms) between the observed and the model predicted positions of the multiple images, defined as
\begin{equation}
    \label{eq:rms}  
    \mathrm{rms} = \sqrt{\frac{1}{N}\sum_{i=1}^N\left|\bm{\theta}_i^\mathrm{obs}-\bm{\theta}_i^\mathrm{pred}\right|^2} \, ,
\end{equation}
(4) the Bayesian Information Criterion \citep[BIC,][]{Schwarz1978}, given by 
    \begin{equation}
        \mathrm{BIC} = k \ln{N_{\rm obs}} + \chi^2 \, ,
    \end{equation}
where $k$ is the number of free parameters and $N_{\rm obs}$ is the number of data points ($=2N$, for the $x$ and $y$ coordinates of the multiple images), and (5) the corrected Akaike Information Criterion \citep[AICc,][]{Akaike1974, CAVANAUGH1997201}, defined as 
    \begin{equation}
        \mathrm{AICc} = 2  k + \chi^2 + \frac{2k(k+1)}{N_{\rm obs}-k-1} \, .
    \end{equation}
The BIC and AICc estimators penalize models with an increasing number of free parameters, to contrast overfitting. Thus, models with lower BIC and AICc (as well as rms and minimum $\chi^2$) values are preferred. 

Additionally, we estimated median values and uncertainties for the model parameters from MCMCs of $10^7$ steps, with acceptance rates between 20\% and 30\%, and rejecting the first 10\% burn-in steps. These chains are the final step of a sequence in which each intermediate chain is used to estimate the covariance matrix of the model parameters and to extract the starting point for the following one. To obtain sensible confidence intervals for the values of the parameters from the chains, we rescaled the adopted positional uncertainties until the $\chi^2$ value was similar to the number of degrees of freedom.

Thanks to the relatively simple structure of group-scale systems, we described the total mass distribution of SDSS~J0100+1818 and disentangled the different mass contributions by assuming a straightforward composite model, with different mass components describing those of the BGG, of the group members, and of the extended DM halo. In the following subsections, we will detail the different parametrizations considered, that will give rise to the four models presented in Section~\ref{sec:lensing_results}.

\subsection{Inclusion of group members}
\label{sec:lensing_group_members}
We modeled the dimensionless surface mass density (convergence, $\kappa$) of each group member as a dual pseudoisothermal elliptical mass distribution \citep[dPIE;][]{Eliasdottir2007}. In \texttt{GLEE}, this profile is described by six parameters: the $x$ and $y$ coordinates of the center, the semiminor ($b$) to semimajor ($a$) axis ratio $q=b/a$, the position angle $\theta$ (measured counter clockwise from $+x$), the Einsten radius $\theta_\mathrm{E}$, the core radius $r_\mathrm{core}$, and the truncation radius $r_\mathrm{T}$. Through this work, we assumed the total mass distribution of the members to be spherical ($q=1$), and without a core radius ($r_\mathrm{core} = 0$). Within these approximations, the convergence becomes 
\begin{equation}
\label{eq:kdpie}
    \kappa_\mathrm{dPIE}(x,y) = \frac{\theta_\mathrm{E}}{2} \left( \frac{1}{R} - \frac{1}{\sqrt{R^2+r_\mathrm{T}^2}} \right) \, ,
\end{equation}
where $R=\sqrt{x^2+y^2}$, and 
\begin{equation}
\label{eq:thetaE}
    \theta_\mathrm{E} = \frac{4 \pi \sigma^2}{c^2} \, ,
\end{equation}
with $\sigma$ representing the effective velocity dispersion of the deflector.
In our analysis, we imposed the total mass of each group galaxy based on its luminosity, $L$, as it is commonly done in cluster-scale strong lensing modeling \citep[e.g.,][]{Grillo2015, Caminha2019, Acebron2022, Bergamini2023}. We estimated $L$ from the Kron magnitude measured with {\tt SExtractor} in the HST F160W band, and we linked the $i$-th galaxy total mass, described by the values of the Einstein radius, $\theta_{{\rm E},\, i}$ and truncation radius, $r_{{\rm T},\,i}$, and its luminosity via the following scaling relations:
\begin{equation}
    \theta_{{\rm E},\, i} = \theta_{{\rm E},\, {\rm ref}}  \left( \frac{L_i}{L_{\rm ref}}\right)^{\alpha} \, \, \mathrm{and}  \, \, \,
r_{{\rm T},\,i} = r_{{\rm T},\,{\rm ref}} \left( \frac{L_i}{L_{\rm ref}}\right)^{\beta} \, \, ,
\label{eq:scaling_reln}
\end{equation}
where $\theta_{{\rm E},\, {\rm ref}}$ and $r_{{\rm T},\,{\rm ref}}$ are reference values relative to a galaxy at the redshift of the group. In particular, we selected the BGG as the reference galaxy, when it is included in the scaling relations, and the second brightest group member at (RA, dec) = (1:00:48.68, +18:18:12.67) ($m = 19.94$ mag), when the BGG is parametrized separately, i.e. without its parameter values following the scaling relations. In this latter case, the models will be labeled with "{\tt +BGG}" at the end of the name. \\
We fixed $\alpha=0.7$ and $\beta=0.5$ \citep[e.g.,][]{Grillo2015, Wang2022}. Given that for a dPIE profile the total mass, $M_{\rm T}$, is proportional to $\theta_\mathrm{E}$ and $r_\mathrm{T}$, the total mass-to-light ratio, for the assumed values of $\alpha$ and $\beta$, increases with the luminosity as
\begin{equation}
    \frac{M_{\rm T}}{L} \sim \frac{\theta_\mathrm{E} r_\mathrm{T}}{L} \sim \frac{L^{0.7} L^{0.5}}{L} \sim L^{0.2} \, ,
\end{equation}
corresponding to the ``tilt of the Fundamental Plane'' observed in elliptical galaxies \citep{Faber1987, Bender1992, Ciotti1996, Bernardi2003, GrilloGobat2010}. 

\subsection{Dark matter halo}
We completed the total mass modeling of the group with the inclusion of an additional mass component, representing, in particular, the group-scale DM halo and all the contributions not associated to the member galaxies. We explored two different mass profiles describing the extended and smooth mass distribution of the group. The first one is a pseudo-isothermal elliptical mass distribution \citep[PIEMD;][]{Kneib1996}. In {\tt GLEE}, it is defined by six parameters: the $x$ and $y$ coordinates of the center, the semi-minor ($b$) to semi-major ($a$) axis ratio $q=b/a$, the position angle $\theta$ (measured counter clockwise from $+x$), the Einstein radius $\theta_\mathrm{E}$, and the core radius $r_\mathrm{core}$. We stress that the value of $\theta_\mathrm{E}$ is defined for a source at $z = \infty$ and does not correspond to that of the Einstein radius of the system, which should be nearly independent of the mass modeling details. The value of $\theta_\mathrm{E}$ is a parameter which describes the lens strength and enters the dimensionless surface mass density $\kappa_\mathrm{PIEMD}$ as
\begin{equation}
\label{eq:kpiemd}
    \kappa_\mathrm{PIEMD} (x,y) = \frac{\theta_\mathrm{E}}{2 \sqrt{r_\mathrm{core}^2+\frac{x^2}{(1+e)^2}+\frac{y^2}{(1-e)^2}}} \, ,
\end{equation}
where the ellipticity $e=\frac{1+q}{1-q}$. 
The second profile is a singular power law elliptical mass distribution \citep[SPEMD;][]{Barkana1998}. In {\tt GLEE}, it is characterized by seven parameters: the first six are in common with the PIEMD profile, and the seventh is the slope $g$, which is related to the three-dimensional logarithmic mass density slope $\gamma'= -\diff \log[\rho(r)]/\diff \log(r)$ (i.e., $\rho \propto r^{-\gamma'}$) through $\gamma'=2 g +1$ (i.e., an isothermal profile corresponds to $\gamma'=2$ and $g = 0.5$). In the following, we will refer to the physical parameter $\gamma'$.
Similarly to the PIEMD case, $\theta_\mathrm{E}$ is a parameter of the mass distribution introduced in the dimensionless surface mass density $\kappa_\mathrm{SPEMD}$ as
\begin{equation}
\label{eq:kspemd}
    \kappa_\mathrm{SPEMD}(x,y) = \theta_\mathrm{E} \left( x^2+\frac{y^2}{q^2} + \frac{4r^2_\mathrm{core}}{(1+q)^2}\right) ^ {-\frac{\gamma^\prime-1}{2}} \, .
\end{equation}
Following the definitions implemented in \texttt{GLEE}, the values of the $\theta_\mathrm{E}$ parameters adopted in the PIEMD and in the SPEMD cases differ by a factor of ($1+q$). 
%\shs{Just to double check, you do *NOT* have the "scale:" keyword included in the theta\_e parameter for the piemd and spemd profiles, right?}

\subsection{Mass models and results}
\label{sec:lensing_results}

We tested four different mass models, exploring two alternative profiles for the smooth, extended group mass distribution (PIEMD and SPEMD) and two options for the BGG in terms of scaling relations. Given the nature of the group, with the BGG being almost three magnitudes brighter than the second brightest galaxy and likely including a large fraction of the total mass of the group members, we tried two different options for the BGG: one in which the BGG follows the scaling relations of Eq.~(\ref{eq:scaling_reln}), and another one in which the total mass of the BGG is still parametrized by a dPIE profile, but independently of the other members. This second option allows the model to associate with more freedom the total mass contribution related to the BGG and to the other group members. To employ the lowest number of free parameters, we fixed the BGG position to its luminosity center, $q=1$, and $r_{\rm core}=0$, leaving its total mass described only by $\theta_\mathrm{E}$ and $r_\mathrm{T}$. The four considered models are thus:

$\bullet$ \piemd: This model includes a PIEMD mass distribution for the extended halo, and dPIEs for all the group members. The BGG is included in the scaling relations, so that the total mass of each group member is related to that of the BGG. This mass model is described by 8 free parameters: 6 relative to the extended halo and 2 to the scaling relations. Given that for all the models the number of observables is equal to 36 ($x$ and $y$ coordinates of each of the 18 total multiple images), this model has $\mathrm{ndof}=18$. 

$\bullet$ \spemd: Similar to the \piemd\ model, this model includes a SPEMD mass distribution for the extended halo, and dPIEs for the group members, including the BGG. It is described by 9 free parameters (adding $\gamma'$), and has $\mathrm{ndof}=17$.

$\bullet$ \piemdbgg: This model includes a PIEMD mass distribution for the extended halo, an independent dPIE for the BGG, and dPIEs for the group members, linked through the scaling relations. This mass model is described by 10 free parameters: 6 relative to the extended halo, 2 relative to the dPIE of the BGG, and 2 to the scaling relations, and has $\mathrm{ndof}=16$.

$\bullet$ \spemdbgg: This model considers a SPEMD mass distribution for the extended halo, an independent dPIE for the BGG, and dPIEs for the group members linked through the scaling relations. It is described by 11 free parameters and has $\mathrm{ndof}=15$.

The best-fit values of the parameters corresponding to each model are reported in Table~\ref{tab:best-fit}. By comparing the best-fit parameter values of these different models, we noticed that the center of the total mass approximately coincides with the luminosity center, when the BGG is modeled inside the scaling relations, while it is shifted by $\approx 0.2\arcsec$ in the $-x$-direction, when it is not. The value of the axis ratio $q$ of the extended halo does not vary significantly among all the different models, as well as that of the position angle $\theta$. The \piemd\ and \spemd\ models predict similar values of the Einstein radii, which describe the strength of the lens, both for the extended halo and the BGG ($\theta_{{\rm E},\, {\rm ref}}$). This latter value of approximately $4.7\arcsec$ is found also when the BGG is modeled outside of the scaling relations (in the \piemdbgg\ and \spemdbgg\ models), thus strengthening its robustness. The core radius of the diffuse halo increases by $\approx 15\%$, when it is parametrized with a SPEMD mass distribution, whose best-fit slope ($\gamma' \sim 2.2$) is slightly steeper than isothermal. The BGG has a truncation radius ($r_{{\rm T},\, {\rm ref}}$ in \piemd\ and \spemd, and $r_{\rm T}$ in \piemdbgg\ and \spemdbgg) larger than the distance over which the multiple images appear (see Table~\ref{tab:multiple_images}).

Summarizing, we have found that the best-fit values of the parameters are stable when the BGG is modeled outside of the scaling relations. We note that, by reducing the number of degrees of freedom from 18 of the \piemd\ model to 15 of the \spemdbgg\ one, the $\chi^2_{\rm min}$ value decreases from 86.63 to 76.82, with a corresponding reduction of the rms value from 0.23\arcsec\ to 0.21\arcsec.

\begin{table}[]
\caption{Results obtained for the deflector total mass distribution from the four adopted models. }
\centering
\scriptsize
\begin{tabular}{@{}lccccccc@{}}
\toprule
\toprule
\multicolumn{2}{c}{\normalsize \piemd } & & & & & &  \\ 
$x$ [$''$] & $y$ [$''$] & $q$ & $\theta$ [rad] & $\theta_\mathrm{E}$ [$''$] &  $r_{\mathrm{core}}$ [$''$]  &  \\  
$-0.01$ & $0.03$ & $0.69$ & $-0.02$ & $20.35$ &  $7.19$  &  \\ 
Scaling rel. & $\theta_{\mathrm{E,\, ref}}$ [$''$] & $r_{\mathrm{T,\,ref}}$ [$''$] & & & & & \\
 & $4.67$ & $11.27$ & & & & & \\
&\multicolumn{2}{c}{\cellcolor{bkg}$N_{\rm obs}=36$}& \multicolumn{2}{c}{\cellcolor{bkg}ndof $=18$} & \multicolumn{2}{c}{\cellcolor{bkg}$\chi^2_{\rm min} = 86.63$} & \\
&\multicolumn{2}{c}{\cellcolor{bkg}$\mathrm{BIC}=151.13$}& \multicolumn{2}{c}{\cellcolor{bkg}$\mathrm{AICc}=162.87$} & \multicolumn{2}{c}{\cellcolor{bkg}$\mathrm{rms}=0.23\arcsec$} & \\
\midrule
\multicolumn{2}{c}{\normalsize \spemd } & & & & & & \\
$x$ [$''$] & $y$ [$''$] & $q$ & $\theta$ [rad] & $\theta_\mathrm{E}$ [$''$] &  $r_\mathrm{core}$ [$''$]  & $\gamma'$ \\
$0.05$ & $0.05$ & $0.70$ & $-0.01$ & $20.53$ &  $8.36$  & $2.18$ \\ 
Scaling rel. & $\theta_{\mathrm{E,\, ref}}$ [$''$] & $r_{\mathrm{T,\,ref}}$ [$''$] & & & & & \\
 & 4.71 & $11.28$ & & & & & \\
&\multicolumn{2}{c}{\cellcolor{bkg}$N_{\rm obs}=36$}& \multicolumn{2}{c}{\cellcolor{bkg}ndof $=17$} & \multicolumn{2}{c}{\cellcolor{bkg}$\chi^2_{\rm min} = 86.62$} & \\
&\multicolumn{2}{c}{\cellcolor{bkg}$\mathrm{BIC}=154.71$}& \multicolumn{2}{c}{\cellcolor{bkg}$\mathrm{AICc}=172.12$} & \multicolumn{2}{c}{\cellcolor{bkg}$\mathrm{rms}=0.23\arcsec$} & \\
\midrule
\multicolumn{2}{c}{\normalsize \piemdbgg } & & & & & &  \\ 
$x$ [$''$] & $y$ [$''$] & $q$ & $\theta$ [rad] & $\theta_\mathrm{E}$ [$''$] &  $r_\mathrm{core}$ [$''$]  &  \\  
$-0.20$ & $0.04$ & $0.68$ & $-0.02$ & $19.50$ &  $7.23$  &  \\ 
 dPIE BGG & $\theta_\mathrm{E}$ [$''$] & $r_{\mathrm{T}}$ [$''$] & & & & & \\
  & $4.65$ & $17.94$ & & & & & \\
Scaling rel. & $\theta_{\mathrm{E,\, ref}}$ [$''$] & $r_{\mathrm{T,\,ref}}$ [$''$] & & & & & \\
 & $2.74$ & $0.65$ & & & & & \\
&\multicolumn{2}{c}{\cellcolor{bkg}$N_{\rm obs}=36$}& \multicolumn{2}{c}{\cellcolor{bkg}ndof $=16$} & \multicolumn{2}{c}{\cellcolor{bkg}$\chi^2_{\rm min} = 76.85$} & \\
&\multicolumn{2}{c}{\cellcolor{bkg}$\mathrm{BIC}=148.52$}& \multicolumn{2}{c}{\cellcolor{bkg}$\mathrm{AICc}=172.85$} & \multicolumn{2}{c}{\cellcolor{bkg}$\mathrm{rms}=0.21\arcsec$} & \\
\midrule
\multicolumn{2}{c}{\normalsize \spemdbgg} & & & & & & \\
$x$ [$''$] & $y$ [$''$] & $q$ & $\theta$ [rad] & $\theta_\mathrm{E}$ [$''$] &  $r_\mathrm{core}$ [$''$]  & $\gamma'$ \\
$-0.19$ & $0.05$ & $0.68$ & $-0.02$ & $19.45$ &  8.47  & $2.17$ \\ 
 dPIE BGG & $\theta_\mathrm{E}$ [$''$] & $r_{\mathrm{T}}$ [$''$] & & & & & \\
  & 4.71 & $18.05$ & & & & & \\
Scaling rel. & $\theta_{\mathrm{E,\, ref}}$ [$''$] & $r_{\mathrm{T,\,ref}}$ [$''$] & & & & & \\
 & 2.69 & $0.69$ & & & & & \\
&\multicolumn{2}{c}{\cellcolor{bkg}$N_{\rm obs}=36$}& \multicolumn{2}{c}{\cellcolor{bkg}ndof $=15$} & \multicolumn{2}{c}{\cellcolor{bkg}$\chi^2_{\rm min} = 76.82$} & \\
&\multicolumn{2}{c}{\cellcolor{bkg}$\mathrm{BIC}=152.07$}& \multicolumn{2}{c}{\cellcolor{bkg}$\mathrm{AICc}=184.82$} & \multicolumn{2}{c}{\cellcolor{bkg}$\mathrm{rms}=0.21\arcsec$} & \\
\bottomrule \\
\end{tabular}
\tablefoot{For each model, we show the best-fit values of the parameters, the number of degrees of freedom (ndof), the minimum chi-square ($\chi^2_{\rm min}$), and the rms values. 
The values of the $x$ and $y$ coordinates are referred to the center of light of the main elliptical lens galaxy, i.e., to its brightest pixel. The position angle $\theta$ is measured counter clockwise from the $x$-axis, aligned following the horizontal direction of Fig.~\ref{fig:labels}. We report in the table only the values of the optimized parameters, while for each group member we fix its total mass center to its luminosity center ($\equiv(0,0)$ for the BGG), and the axis ratio and core radius values to $q=1$ and $r_{\rm core}=0$, respectively.}
\label{tab:best-fit}
\end{table}

\subsection{Mass profiles of the deflectors}
\begin{figure*}
   \centering
   \includegraphics[width=0.95\textwidth]{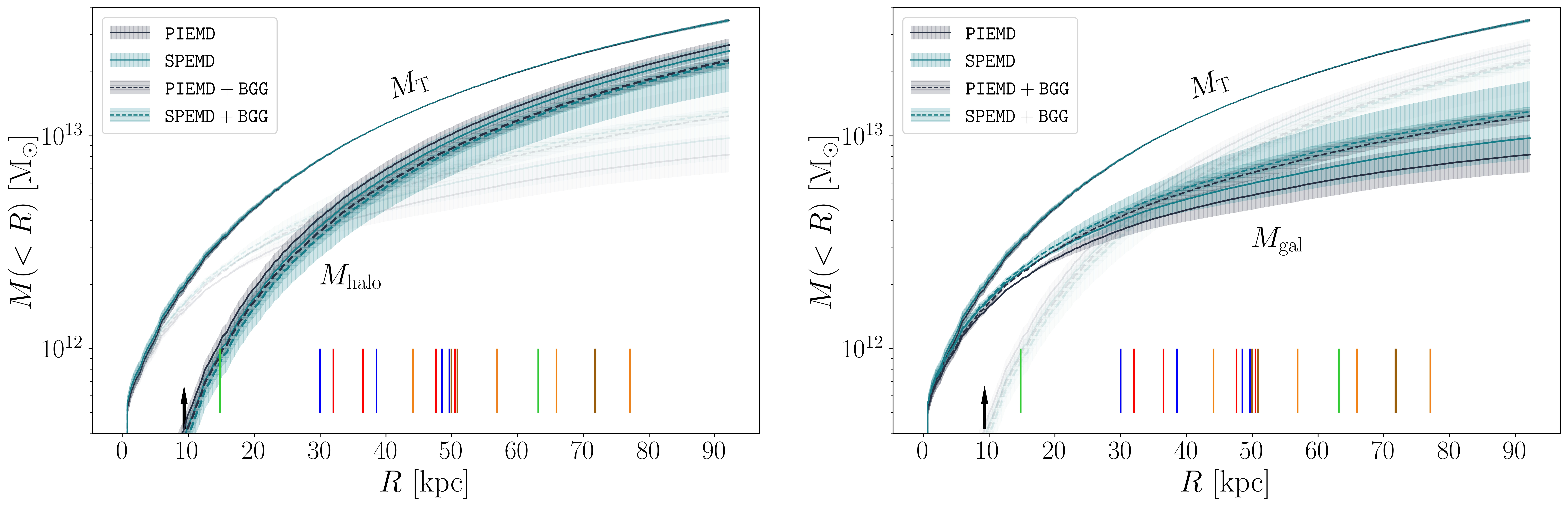}
      \caption{Cumulative projected total mass profiles for the \piemd\ (solid black), \spemd\ (solid blue), \piemdbgg\ (dashed black) and \spemdbgg\ (dashed blue) models with $\pm$1$\sigma$ uncertainties (differently hatched shaded areas). 
      The curves on top, labeled with $M_\mathrm{T}$, are relative to the total mass, and they are divided into the halo component ($M_\mathrm{halo}$, left panel) and the galaxy member component ($M_\mathrm{gal}$, right panel), identified by the same linestyle. For comparison, we leave $M_\mathrm{gal}$ and $M_\mathrm{halo}$ as faint lines in the left and right panels, respectively.
      The vertical lines close to the $x$-axis locate the distances from the lens galaxy center of the different multiple images, color-coded following Fig.~\ref{fig:labels}. The black arrow at $R=9.32$~kpc shows the effective radius of the BGG. }
    \label{fig:mass_halo_galaxies}
   \end{figure*}

We measured the cumulative total mass ($M_{\rm T}$) profile for each model by randomly extracting 1000 parameter value sets from the last MCMC chains of $10^7$ steps described above. \texttt{GLEE} can create the convergence maps, that we converted into total mass maps. Then, we summed up the contribution of all the pixels within circular apertures centered on the brightest pixel of the main lens galaxy, with a step of $0.5$ pixels, to obtain the cumulative total mass profiles that are presented in the following. For each aperture, we estimated the 1$\sigma$ uncertainties as the $16^\mathrm{th}$ and $84^\mathrm{th}$ percentile values of the distribution of the total mass values of all the 1000 random models. We separated the total mass contribution of the extended halo (PIEMD or SPEMD) from that of the group members (dPIE), by considering the relevant free parameters in the MCMC chains. The resulting cumulative mass profiles are shown in Fig.~\ref{fig:mass_halo_galaxies}. The four models predict fully consistent total mass profiles, with mean relative uncertainties of only 2\%. The total mass value projected within 50~kpc, approximately equivalent to the Einstein radius of the system, is $(1.55 \pm 0.01) \times 10^{13}$~M$_\odot$, while within the projected distance of the furthest multiple image ($R \approx 77$~kpc) is $(2.78 \pm 0.04) \times 10^{13}$~M$_\odot$. 

The mass component associated with the diffuse halo presents a sligthly larger mean uncertainty of around 10\% in all the four models. For this component, the models predict a projected total mass value of $9.2^{+1.2}_{-3.1} \times 10^{12}$~M$_\odot$, at $R=50$~kpc, and of $1.9^{+0.2}_{-0.7} \times 10^{13}$~M$_\odot$, at $R=77$~kpc. As visible in Fig.~\ref{fig:mass_halo_galaxies}, the profiles show that, in the central regions, the total mass is dominated by the group member contribution, while from $R \approx 35$~kpc outwards the extended halo prevails. The different models predict consistent relative weights for the different mass components, with the models with the BGG outside of the scaling relations assigning to the extended halo a slightly lower contribution.  

\begin{figure}
   \centering
   \includegraphics[width=\columnwidth]{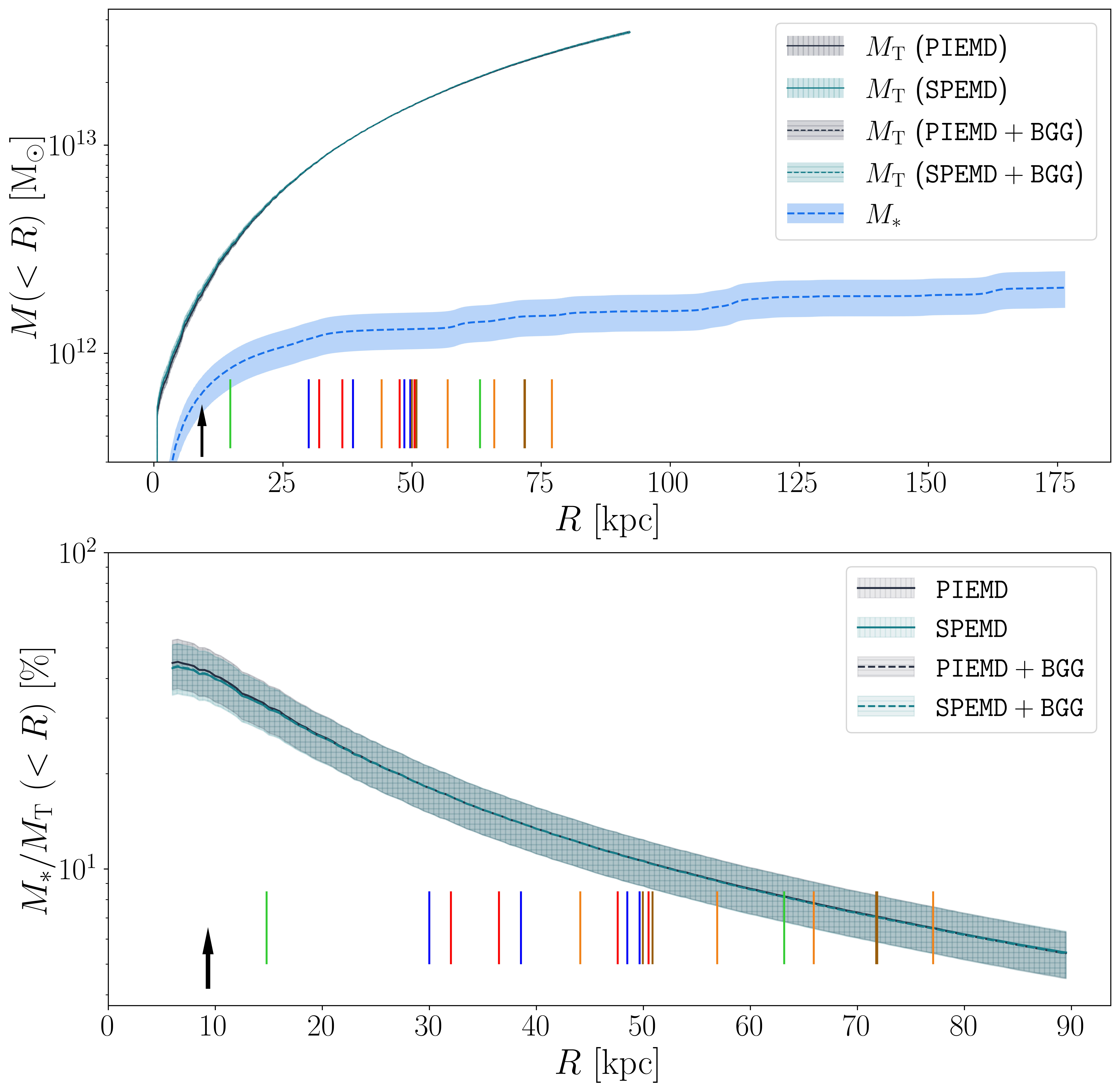}
      \caption{Top: cumulative projected total mass profiles for the \piemd\ (solid black), \spemd\ (solid blue), \piemdbgg\ (dashed black) and \spemdbgg\ (dashed blue) models with $\pm$1$\sigma$ uncertainties (differently hatched shaded areas), as in Fig.~\ref{fig:mass_halo_galaxies}, compared with the cumulative projected stellar mass profile with $\pm$1$\sigma$ uncertainties, represented by the light blue dashed curve and shaded area, respectively. 
      Bottom: cumulative projected stellar-over-total mass fraction profiles, color coded for each model according to the top panel. 
      In both panels, the vertical lines close to the $x$-axis locate the distances from the lens galaxy center of the different multiple images, color-coded following Fig.~\ref{fig:labels}. The black arrow at $R=9.32$~kpc shows the effective radius of the BGG. }
   \label{fig:stellar_mass_profile}
   \end{figure}

We measured the stellar mass ($M_*$) profile of the BGG in \citetalias{Bolamperti2023} from its reconstructed luminosity profile, by assuming a constant stellar mass-to-light ratio estimated through a spectral energy distribution (SED) fitting. We assume here the same value for the stellar mass-to-light ratio of the other galaxy members and convert the HST F160W image into a $M_*$ map, by multiplying it by the BGG mass-to-light ratio and by a mask designed to include only the pixels associated to the 19 group members (including the BGG). This is justified by the fact that all the group members show similar spectra and colors in the MUSE and photometric data, and thus they are likely early-type galaxies hosting similar stellar populations.
From the resulting image, we measured the cumulative projected stellar mass profile of the group by summing the contribution of the pixels within circular apertures centered on the brightest pixel of the main lens galaxy, with a step of $0.5$ pixels. The profile is shown in the top panel of Fig.~\ref{fig:stellar_mass_profile}. The BGG clearly dominates the light (and thus the stellar mass) distribution, and the inclusion in the external apertures of the other group members is visible as steps in the profile. Given that we can measure the stellar mass distribution of the group all across the MUSE field of view, it is possible to explore the outer regions, up to $R \approx 30\arcsec \approx 180$ kpc. We note that the total stellar mass value of the group at $R = 180$ kpc is approximately equal to $1.5$ times that of the BGG and, at the largest distance where the multiple images appear ($R\approx 77$~kpc), the total stellar mass of the group is $(1.7 \pm 0.3) \times 10^{12}$ M$_\odot$, and the BGG contributes with almost 90\% of it, namely with $(1.5 \pm 0.3) \times 10^{12}$ M$_\odot$. We remark that the cumulative stellar mass profile is between 1.5 and 5 times smaller than the total one associated with the galaxy member components, shown in the right panel of Fig.~\ref{fig:mass_halo_galaxies}. This difference is mainly due to the presence of DM, in the form of galaxy-scale halos, in the group members.

In the bottom panel of Fig.~\ref{fig:stellar_mass_profile}, we plot the cumulative projected stellar-over-total mass fraction profiles relative to each model. They are all in very good agreement, differing only by less than 2\% in the inner region, and by a fraction of percent from $R>20$~kpc outwards. We measured a stellar-over-total mass fraction value of $45.6^{+8.7}_{-8.3} \%$ at the lens galaxy effective radius ($R_{\rm e} = 9.32$~kpc), decreasing to $(11.1\pm 1.9) \%$ at $R=50$~kpc and to $(6.6\pm 1.1) \%$ at $R \approx 77$~kpc.  
%\shs{How do the M* profile compare to the M profile of the galaxy-member component in Fig.7?  It looks like Fig.7 is substantially higher by a factor of a few.  Is it due to presence of gas, and additional DM associated with the cluster-member components in Fig.7?  }

\section{Strong lensing models with variable cosmology}
\label{sec:lensing_model_cosmo}
\subsection{Single-plane models}
The observed positions of the lensed multiple images of a system depend both on the total mass distribution of the lens and on the cosmological angular-diameter distances between the observer, the lens, and the sources. Systems where multiple sources at different redshifts are strongly lensed by the same foreground deflector allow one to break the mass-cosmology degeneracy, and thus to learn about the geometry of the Universe. By measuring the ratio of different Einstein radii in a multiple-source system, and assuming a cosmological model, it is possible to estimate the values of the cosmological parameters defining the global geometry of the Universe, without any dependence on the value of $H_0$ \citep{Collett2012}. The same parameter values can be measured thanks to other cosmological probes  \citep{Eisenstein2005, Komatsu2014, Treu2022}.
In light of the results of the statistical estimators adopted and discussed in the previous sections, we decided to consider in the following only the \piemd\ and \spemd\ models. We assumed a flat ($\Omega_\Lambda + \Omega_{\rm m} = 1 $) $\Lambda$CDM cosmology and reoptimized the \piemdOm\ and \spemdOm\ models, where $\Omega_{\rm m}$ was introduced as a free parameter. We further reconsidered the \piemdOmw\ and \spemdOmw\ models in a flat $w$CDM cosmology, where both $\Omega_{\rm m}$ and $w$ were free to vary. We reran the same procedure described above, starting from a $\chi^2$ minimization on the lens plane and then obtaining a sequence of MCMC chains, resulting in a final one composed of $10^7$ steps. We adopted uniform priors on the values of $\Omega_{\rm m}$ and $w$, respectively, between 0 and 1, and $-2$ and 0. 

To fully exploit our observational dataset and alleviate possible degeneracies bewteen the values of the cosmological and lens total mass distribution parameters, we used the available kinematic information on the BGG to impose a Gaussian prior on the value of its Einstein radius $\theta_{\mathrm{E,\, ref}}$. As we wished to probe the inner kinematics of the BGG, we weighed the MUSE cube with its observed surface brightness in the HST F160W band, degraded and re-binned to the PSF and pixel-scale of MUSE. We then extracted a spectrum within an aperture with a radius of $3\arcsec$, centered on the BGG. This light-weighed extraction procedure yields measured velocity dispersion values equivalent to those that would be obtained from a non-weighed spectrum extracted within the galaxy effective radius (Granata et al. in prep.), whilst increasing the $S/N$ of the extracted spectrum compared to a small straightforward aperture. By doing so, we measured a stellar velocity dispersion value of $(380.5 \pm 4.4)\, \si{km.s^{-1}}$. We converted this value, with a conservative uncertainty of $10\, \si{km.s^{-1}}$, to the corresponding Einstein radius for an isothermal profile, through Eq.~\ref{eq:thetaE}. We note that the value obtained in this way is very similar to the best-fit values of $\theta_{\mathrm{E,\, ref}}$ and $\theta_{\rm E}$ of the BGG reported in Table~\ref{tab:best-fit}, as well as consistent within about 1$\sigma$ with the results from the respective posterior probability distributions (e.g., $(409 \pm 29) \, \si{km.s^{-1}}$ for the \piemdbgg\ model).
Furthermore, we checked that the introduction of this prior does not have any significant impact on the lens mass profiles reconstructed through the different models. In detail, all the profiles shown in Figs.~\ref{fig:mass_halo_galaxies} and \ref{fig:stellar_mass_profile} did not vary appreciably. %\shs{The use of kinematics would also help study the cluster mass profile in Section 4, not only cosmology in Section 5.  How about incorporating the kinematic constraint also in the results of Section 4?   Otherwise, it seems odd that we use kinematic info for Section 5, but not Section 4, unless there are good reasons for this difference in data.}

\begin{figure}
   \centering
   \includegraphics[width=0.9\columnwidth]{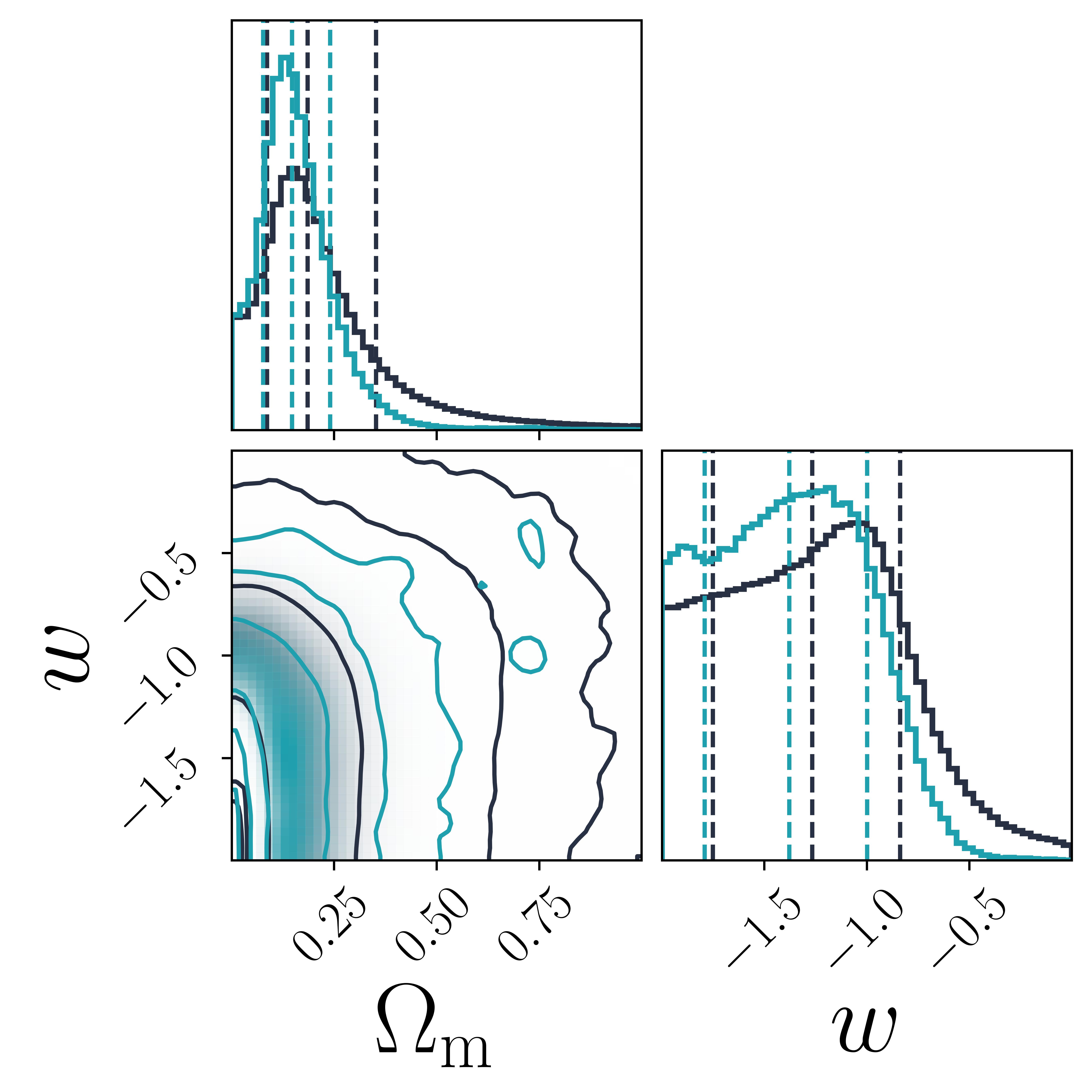}
      \caption{Probability density distributions of the cosmological parameters $\Omega_{\rm m}$ and $w$ in the \piemdOmw\ (black) and \spemdOmw\ (blue) models. We show the marginalized 1D histograms of each parameter and their joint 2D probability distribution. The vertical dashed lines in the 1D histograms represent the 16th, 50th, and 84th percentiles, while the solid lines in the 2D distributions represent the 0.68, 0.95 and 0.99 contour levels.}
   \label{fig:cosmo_corner}
   \end{figure}

We report in Table~\ref{tab:cosmo} the median values, with the 68\% confidence level (CL) uncertainties, of the cosmological parameters $\Omega_{\rm m}$ and $w$, and show in Fig.~\ref{fig:cosmo_corner} their posterior probability distributions for the \piemdOmw\ and \spemdOmw\ models. The average 1$\sigma$ uncertainty is on the order of 0.1 for $\Omega_{\rm m}$ and 0.4 for $w$. The $\Omega_{\rm m}$ distributions are skewed towards the lower limit of our uniform prior (i.e., zero), thus in the table we also include the 95.4\% quantile upper limits. The models with $w$ as a free parameter predict median values of $w$ smaller than $-1$, with relative uncertainties of approximately $30\%$. In particular, we found $w = -1.27_{-0.48}^{+0.43}$ and $w = -1.38_{-0.41}^{+0.38}$ for the \piemdOmw\ and \spemdOmw\ models, respectively. Given that the $w$ distributions decrease quite slowly towards small $w$ values, the resulting median values and lower limits are sensitive to the extent of the (flat) prior. We remark that our results were obtained with a flat and quite large prior (i.e., $w \in [-2,0]$). We additionally report here the more robust estimates of the 95.4\% quantile upper limits for $w$, of $-0.53$ and $-0.79$ for the \piemdOmw\ and \spemdOmw\ models, respectively. The considered statistical estimators did not reveal a strongly preferred model, as all of them reproduced the observed positions of the multiple images with the same $\mathrm{rms}$ of $0.23\arcsec$ (and thus had very similar $\chi^2_{\rm min}$ values). Consequently, the \piemdOm\ model, given its smallest number of free parameters, is moderately preferred by both the BIC and AICc metrics, followed by the \piemdOmw\ model. 

\begin{table}[]
\caption{Cosmological parameters $\Omega_{\rm m}$ and $w$ from the six adopted strong lensing models with variable cosmology.}
\centering
\footnotesize
\begin{tabular}{@{}lccccccc@{}}
\toprule
\toprule
\multicolumn{2}{c}{Model} & \multicolumn{2}{c}{$\Omega_{\rm m}$}  & \multicolumn{2}{c}{$\Omega_{\rm m}$} & \multicolumn{2}{c}{$w$}  \\ 
\multicolumn{2}{c}{} & \multicolumn{2}{c}{}  & \multicolumn{2}{c}{95.4\% upper limit} & \multicolumn{2}{c}{}  \\ 
\midrule
\multicolumn{3}{l}{\normalsize \piemdOm }  & & & & &  \\ 
\multicolumn{2}{c}{} & \multicolumn{2}{c}{$0.14_{-0.09}^{+0.16}$}  & \multicolumn{2}{c}{$0.47$} & \multicolumn{2}{c}{} \\ 
&  & \multicolumn{2}{c}{\cellcolor{bkg}$N_{\rm obs}=36$}& \multicolumn{2}{c}{\cellcolor{bkg}ndof $=17$} & \multicolumn{2}{c}{\cellcolor{bkg}$\chi^2_{\rm min} = 85.95$} \\
& & \multicolumn{2}{c}{\cellcolor{bkg}$\mathrm{BIC}=154.04$}& \multicolumn{2}{c}{\cellcolor{bkg}$\mathrm{AICc}=171.45$} & \multicolumn{2}{c}{\cellcolor{bkg}$\mathrm{rms}=0.23\arcsec$} \\
\midrule
\multicolumn{3}{l}{\normalsize \spemdOm }  & & & & &  \\ 
\multicolumn{2}{c}{} & \multicolumn{2}{c}{$ 0.09_{-0.06}^{+0.10}$}  & \multicolumn{2}{c}{$0.29$} & \multicolumn{2}{c}{}  \\ 
& &\multicolumn{2}{c}{\cellcolor{bkg}$N_{\rm obs}=36$}& \multicolumn{2}{c}{\cellcolor{bkg}ndof $=16$} & \multicolumn{2}{c}{\cellcolor{bkg}$\chi^2_{\rm min} = 85.94$} \\
& &\multicolumn{2}{c}{\cellcolor{bkg}$\mathrm{BIC}=157.61$}& \multicolumn{2}{c}{\cellcolor{bkg}$\mathrm{AICc}=181.94$} & \multicolumn{2}{c}{\cellcolor{bkg}$\mathrm{rms}=0.23\arcsec$} \\
\midrule
\multicolumn{3}{l}{\normalsize \piemdOmw }  & & & & &  \\ 
 \multicolumn{2}{c}{} & \multicolumn{2}{c}{$0.19_{-0.10}^{+0.17}$}  & \multicolumn{2}{c}{$0.59$} & \multicolumn{2}{c}{$-1.27_{-0.48}^{+0.43}$} \\ 
& &\multicolumn{2}{c}{\cellcolor{bkg}$N_{\rm obs}=36$}& \multicolumn{2}{c}{\cellcolor{bkg}ndof $=16$} & \multicolumn{2}{c}{\cellcolor{bkg}$\chi^2_{\rm min} = 85.68$} \\
& &\multicolumn{2}{c}{\cellcolor{bkg}$\mathrm{BIC}=157.35$}& \multicolumn{2}{c}{\cellcolor{bkg}$\mathrm{AICc}=181.68$} & \multicolumn{2}{c}{\cellcolor{bkg}$\mathrm{rms}=0.23\arcsec$} \\
\midrule
\multicolumn{3}{l}{\normalsize \spemdOmw }  & & & & &  \\ 
\multicolumn{2}{c}{}&\multicolumn{2}{c}{$0.15_{-0.07}^{+0.10}$}  & \multicolumn{2}{c}{$0.34$} & \multicolumn{2}{c}{$-1.38_{-0.41}^{+0.38}$} \\ 
& &\multicolumn{2}{c}{\cellcolor{bkg}$N_{\rm obs}=36$}& \multicolumn{2}{c}{\cellcolor{bkg}ndof $=15$} & \multicolumn{2}{c}{\cellcolor{bkg}$\chi^2_{\rm min} = 85.68$} \\
& &\multicolumn{2}{c}{\cellcolor{bkg}$\mathrm{BIC}=160.93$}& \multicolumn{2}{c}{\cellcolor{bkg}$\mathrm{AICc}=193.68$} & \multicolumn{2}{c}{\cellcolor{bkg}$\mathrm{rms}=0.23\arcsec$} \\
\midrule
\midrule
\multicolumn{3}{l}{\normalsize MP-\piemdOm }  & & & & &  \\ 
 \multicolumn{2}{c}{} & \multicolumn{2}{c}{$0.18_{-0.12}^{+0.29}$}  & \multicolumn{2}{c}{$0.76$} & \multicolumn{2}{c}{} \\ 
& &\multicolumn{2}{c}{\cellcolor{bkg}$N_{\rm obs}=36$}& \multicolumn{2}{c}{\cellcolor{bkg}ndof $=16$} & \multicolumn{2}{c}{\cellcolor{bkg}$\chi^2_{\rm min} = 84.45$} \\
& &\multicolumn{2}{c}{\cellcolor{bkg}$\mathrm{BIC}=156.12$}& \multicolumn{2}{c}{\cellcolor{bkg}$\mathrm{AICc}=180.45$} & \multicolumn{2}{c}{\cellcolor{bkg}$\mathrm{rms}=0.23\arcsec$} \\
\midrule
\multicolumn{3}{l}{\normalsize \mbox{MP-\piemdOmw} }  & & & & &  \\ 
\multicolumn{2}{c}{}&\multicolumn{2}{c}{$0.19_{-0.11}^{+0.30}$}  & \multicolumn{2}{c}{$0.81$} & \multicolumn{2}{c}{$-1.39_{-0.41}^{+0.49}$} \\ 
& &\multicolumn{2}{c}{\cellcolor{bkg}$N_{\rm obs}=36$}& \multicolumn{2}{c}{\cellcolor{bkg}ndof $=15$} & \multicolumn{2}{c}{\cellcolor{bkg}$\chi^2_{\rm min} = 84.42$} \\
& &\multicolumn{2}{c}{\cellcolor{bkg}$\mathrm{BIC}=159.67$}& \multicolumn{2}{c}{\cellcolor{bkg}$\mathrm{AICc}=192.42$} & \multicolumn{2}{c}{\cellcolor{bkg}$\mathrm{rms}=0.23\arcsec$} \\
\bottomrule \\
\end{tabular}
\tablefoot{The median values, with the 68\% CL  uncertainties, of the cosmological parameters $\Omega_{\rm m}$ and $w$ were extracted as the 50$^\mathrm{th}$ quantiles, and from the 16$^\mathrm{th}$ and 84$^\mathrm{th}$ quantiles of the posterior probability distributions from the MCMC chains. Given that the $\Omega_{\rm m}$ distributions are skewed towards values approaching zero, we also include the 95.4\% quantile upper limits. For each model, we also report the adopted metrics describing the goodness of the best-fit models. The two bottom entries are based on the multi-plane approach, with an additional mass distribution at $z=1.880$ (see Sect.~\ref{sec:multi_plane}).}
\label{tab:cosmo}
\end{table}

\subsection{Multi-plane models}
\label{sec:multi_plane}
Given the presence of galaxies (i.e., mass concentrations) at different redshifts, we additionally adopted a multi-plane strong lensing approach \citep{Blandford1986, Schneider1992}, to take into account the fact that the light rays coming from the furthest sources are deflected not only once, by the main group at $z=0.581$, but instead multiple times by other secondary deflectors at different redshifts. In particular, we explored the impact of such an approach on the inferred values of the cosmological parameters. 

In the multi-plane formalism, the total deflection angle is a suitable sum of all the deflection angles relative to the different deflector planes \citep[see, e.g.,][]{Gavazzi2008}. This approach has been used to model both different foreground and background deflectors at distances different from that of the primary lens \citep[e.g.,][Acebron et al., in prep.]{Chirivi2018, Schuldt2024} and to assign mass to multiply imaged background sources, where several sources at different redshifts are lensed by the same deflector \citep[e.g.,][]{Gavazzi2008, Schuldt2019, Collett2020, Wang2022}.

\begin{figure}
   \centering
   \includegraphics[width=0.9\columnwidth]{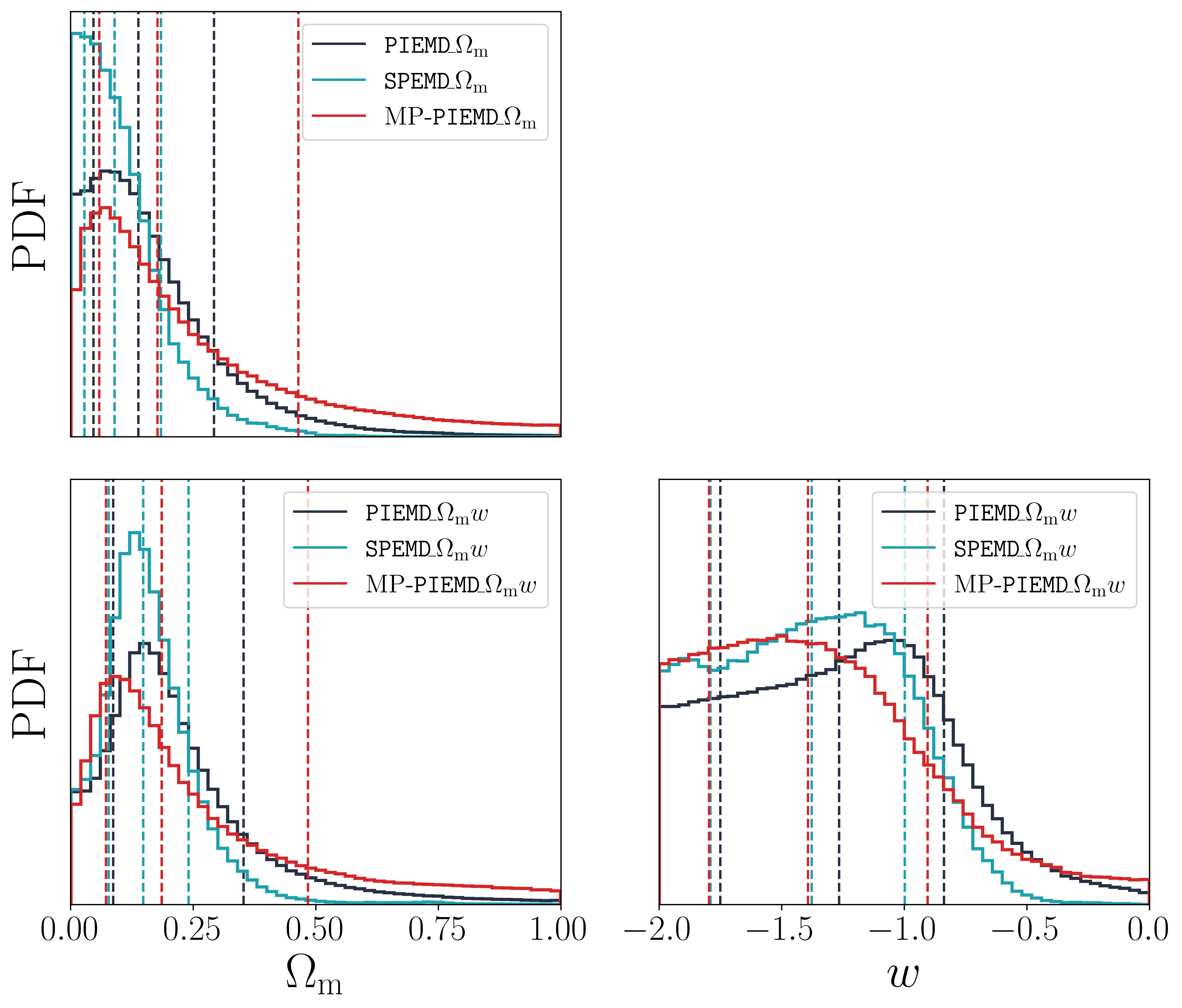}
      \caption{Marginalized 1D probability density distributions of the cosmological parameters $\Omega_{\rm m}$ and $w$ in the \piemdOm\ (black),  \spemdOm\ (blue), and \mbox{MP-\piemdOm} (red) models (top row), and \piemdOmw\ (black),  \spemdOmw\ (blue), and \mbox{MP-\piemdOmw} (red) models (bottom row). The vertical dashed lines represent the 16th, 50th, and 84th percentiles for each distribution.}
   \label{fig:multiplane}
   \end{figure}

The lens models described above predict that the sources A, B, E, and F are located angularly close to the optical axis (i.e., the center of the BGG), while C lies $\sim 2\arcsec$ away. Considering the mutual angular distances, and the pixelated source reconstructions developed in \citetalias{Bolamperti2023}, it is likely that the second largest contribution to the total deflection is that happening at the plane of A and B on the light rays coming from sources E and F. 
Thus, we developed the \mbox{MP-\piemdOm} and \mbox{MP-\piemdOmw} models, that extended the \piemdOm\ and \piemdOmw\ models, respectively, by adding an additional PIEMD mass distribution at the redshift of A and B, $z=1.880$. We imposed it to be spherical ($q=1$) and with a vanishing core radius. Thus, this mass structure is described by three parameters: the coordinates ($x_2$, $y_2$) of the center of mass and $\theta_{\rm E2}$, where the subscripts indicate that these parameters are referred to the secondary lens. We repeated the procedure described above, minimizing the $\chi^2$ on the lens plane, and then running a set of MCMC chains, resulting in a final one of $10^7$ steps. The value of $\theta_{\rm E2}$ was extracted with a uniform large prior, while the values of $x_2$ and $y_2$ were linked to the position of source A (that is brighter and more extended than B). The \mbox{MP-\piemdOm} and \mbox{MP-\piemdOmw} models are able to reproduce the observed positions of the multiple images with rms values of 0.23\arcsec\ (as for the \piemdOm\ and \piemdOmw\ models), but they are penalized by their BIC and AICc scores, due to the inclusion of the additional free parameter (the position of source A was optimized also in the previous models).
From the posterior probability distributions, we observed a clear degeneracy between the parameters describing the centroids of the main halo at $z=0.581$, ($x$, $y$), and of the secondary lens, ($x_2$, $y_2$). 
No clear degeneracy is observed with the cosmological parameters, that present consistent median values and uncertainties similar for $w$ and boosted by a factor of $\sim 1.5$ for $\Omega_{\rm m}$, due to the more extended tails towards large values (see Fig.~\ref{fig:multiplane} and the bottom rows of Table~\ref{tab:cosmo}). We also notice that these extended models predict cumulative total mass profiles for the main deflector that are fully consistent with those shown in Fig.~\ref{fig:mass_halo_galaxies}.

\section{Discussion}
\label{sec:discussion}
\subsection{An enhanced model for SDSS~J0100+1818}
\label{sec:comparison_with_2023}
We compare here our results with those obtained in \citetalias{Bolamperti2023}, highlighting the improvements and the overall consistency. The new MUSE data allowed us to: \\
- spectroscopically confirm the redshift of source C ($z_{\rm C}=1.698$). This value is fully consistent with the best-fit measurements of 1.72 and 1.69 by \citetalias{Bolamperti2023}. This spectroscopic confirmation enabled us to reduce some degeneracies found in the previous work between $z_{\rm C}$ and the parameters (in particular $\theta_\mathrm{E}$ and $\gamma'$) describing the total mass distribution of the deflector; \\
- characterize the deflector plane. We identified 18 group members (excluding the BGG), that were included in the strong lensing model as individual mass components, instead of as an external shear term, and we measured the BGG stellar velocity dispersion profile with an accuracy of $\lesssim 8\%$ (approximately 3\% in the four inner bins); \\
- reduce the relative uncertainties on the cumulative total mass profiles from 17\%, 1\%, and 5\%, to 3\%, $<1$\%, and 1\%, at $R\approx 15$, 42, and 63 kpc, respectively, thanks to the increased number of multiple images from additional sources at different redshifts; \\
- enlarge the radial interval probed by strong gravitational lensing with the discovery of the systems E and F, from the closest multiple image (C2, at $R\approx 15$~kpc) to the furthest (E3 at $R \approx 77$~kpc, previously C1 at $R \approx 63$~kpc). This allowed us to robustly reconstruct the total mass profile at previously unexplored physical distances from the group center with a precision of just a few percent; \\
- confirm the measured median and $\pm 1\sigma$ cumulative total mass profiles (Fig.~\ref{fig:mass_halo_galaxies}), as well as the stellar-over-total mass profiles (Fig.~\ref{fig:stellar_mass_profile}), found (after correction) by \citetalias{Bolamperti2023} over the explored radial distances. In particular, we measure (\citetalias{Bolamperti2023} measured) a projected total mass value of $(1.20 \pm 0.01) \times 10^{13}$~M$_\odot$ ($(1.16 \pm 0.01) \times 10^{13}$~M$_\odot$) within 42~kpc, and a stellar-over-total mass fraction of $45.6^{+8.7}_{-8.3} \%$ at the lens galaxy effective radius, decreasing to $(13.6\pm 2.3) \%$ at $R=42$~kpc ($(38\pm 9) \%$ decreasing to $(12\pm 2) \%$)).

\subsection{SDSS~J0100+1818 as a fossil system}
The hierarchical evolution of structures predicted by the $\Lambda$CDM scenario implies that, in the case of some compact groups of galaxies, all the main galaxies of a group could merge into a single one \citep{Ponman1993}. As this central galaxy cannibalizes more galaxies, the luminosity gap between it and the remaining galaxies increases with time. \cite{Jones2003} firstly gave an operational definition of the so-called ``fossil" groups, namely systems where the magnitude gap in the $r$-band between the first two brightest galaxies, $\Delta m_{12}$, lying at a projected distance smaller than half the value of the virial radius, is larger than 2. Later, \citet{Dariush2010} proposed to consider the magnitude gap between the first and the fourth brightest galaxies, requiring $\Delta m_{14}>2.5$. This criterion was shown to be more robust in classifying the groups with a small number of members, where the serendipitous infall of a bright galaxy would change the fossil status. Simulations show that the fossil system status is a phase in the evolution of a group rather than a class itself, as it can be lost if a bright galaxy is included in the system, but also acquired if the BGG merges with other members. Both definitions also include the presence of a diffuse X-ray emission, with $L_{\rm X} \geq 10^{42}$~erg~s$^{-1}$.
The growing mechanism of fossil groups, given the typical angular momentum loss rates via dynamical friction, occurs over long timescales, and thus the discovery of fossil systems as old, undisturbed groups at intermediate redshift (approximately half of the age of the Universe) can act as a probe for galaxy evolution models.
Moreover, fossil systems have a mass concentration in their center that is higher than that of normal galaxy groups \citep{Johnson2018b}, and this makes them particularly efficient as strong gravitational lenses. 

Given the dominance of the BGG in its center, the SDSS~J0100+1818 group was identified as a candidate fossil system in the CASSOWARY survey, but the lack of observations did not allow one to draw firm conclusions. Even now, we would need additional data to securely assign it the status: future deep X-ray data would be necessary to accurately measure the virial radius and confirm whether SDSS~J0100+1818 is a dynamically evolved system immersed in a fairly undisturbed hot gas halo, as imposed by the criterion. This X-ray condition was originally imposed to guarantee the presence of an extended halo in which the group is immersed, and this is clearly the case from the outcomes of the strong lensing modeling and from the total mass parametrization assumed \citep[see also the studies][about CSWA31, a similar system included in the CASSOWARY survey]{Grillo2013, Wang2022}. Additionally, SDSS~J0100+1818 fulfills both the \citet{Jones2003} and \citet{Dariush2010} criteria, presenting $\Delta m_{12} = 2.89$ and $\Delta m_{14} = 3.35$. 
The confirmation of SDSS~J0100+1818 as a fossil system would be decisive to put new constraints on the mechanisms that drive the formation and evolution of fossil systems and of extremely massive early-type galaxies, given that it would be the most distant system known so far \citep[][and refrences therein]{Khosroshahi2006, Aguerri2021}. SDSS~J0100+1818 redshift is similar to that of the system studied by \citet{Ulmer2005}, which does not satisfies the condition of \citet{Jones2003}, with $\Delta m_{12} = 1.93$. 

\subsection{The values of $\Omega_{\rm m}$ and $w$ from multi-plane strong lensing systems}
%In Section~\ref{sec:lensing_model_cosmo}, we presented our $\Omega_{\rm m}$ and $w$ measurements, obtained thanks to the five different background sources, located in three different redshift planes, strongly lensed by SDSS~J0100+1818. The resulting confidence intervals for $\Omega_{\rm m}$ and $w$ are reported in Table~\ref{tab:cosmo} and shown in Fig.~\ref{fig:cosmo_corner}. \\
We compare here the cosmological measurements obtained in this work with those from the galaxy-scale multi-source-plane strong lensing system SDSS~J0946+1006 \citep[also known as the ``Jackpot'';][]{Gavazzi2008, Smith2021}. The latter is composed of a galaxy at $z_l=0.222$ that acts as a strong lens on three background sources, at redshifts 0.609, 2.035, and 5.975. Even if the Jackpot deflector redshift is lower than that of SDSS~J0100+1818, the lensed sources span very similar redshift intervals, with the $D_\mathrm{ds}/D_\mathrm{s}$ ratios ranging from 0.54 to 0.72 for SDSS~J0100+1818 and from 0.60 to 0.89 for the Jackpot (in the standard cosmological model). As shown in \citet{Collett2012}, a double-source system with the first source at a redshift relatively close to that of the deflector and the second one at a significantly higher redshift is particularly effective in estimating the value of $w$. More in general, different combinations of source redshifts lead to a higher sensitivity in different parts of the $\Omega_{\rm m}$-$w$ plane, making it particularly important in the future to combine the cosmological inferences from a large sample of multi-source strong lensing systems. 
%, with a mild dependency on the redshift of the lens. 
We find that our measurements of $\Omega_{\rm m}$ and $w$ with SDSS~J0100+1818 are comparable, in terms of values and accuracy, to those obtained from the Jackpot in \citet{Collett2020} and \citet{Smith2021}, where they were combined though with priors from \textit{Planck} \citep{Planck2014}.

Given the group nature of SDSS~J0100+1818 and the relatively large number of background sources and their multiple images, we also compare our results with those obtained with lens clusters of galaxies. These systems can rely on a large number of multiple images, from several sources located at different redshifts, that help in breaking the mass-sheet and mass-cosmology degeneracies, making them excellent cosmological probes. On the other hand, cosmographic measurements are hindered by the complex total mass distribution of clusters: in order to accurately reproduce the observed positions of many multiple images, distributed over wide spatial regions, several mass components, tracing both the baryonic (stars and hot-gas) and DM distributions, are usually needed. This complexity can introduce some systematic uncertainties associated to the specific choice of the total mass parametrization.
On the contrary, galaxy and group-scale systems have the advantage that the lens is usually successfully modeled with a few mass components, with an immediate physical interpretation, like the simple DM halo, BGG, and group member contributions in SDSS~J0100+1818. \\
\citet{Caminha2022} exploited five strong lensing clusters (with redshift from 0.234 to 0.587), with 7-20 multiply imaged sources per system, with redshift from 0.68 to 6.85. They combined the results from the lens clusters in the sample and found $\Omega_{\rm m} = 0.30^{+0.09}_{-0.11}$ and $w = -1.12^{+0.17}_{-0.32}$ but, when considering the individual clusters, they obtained 1$\sigma$ uncertainties on $\Omega_{\rm m}$ ranging from 0.2 to 0.4. This relative uncertainty is fully consistent with that obtained from SDSS~J0100+1818, showing that the lower number of multiple images is balanced by the simpler total mass distribution. Moreover, \citet{Caminha2022} also noticed that the typical "L-shape" they observed in the $\Omega_{\rm m}$-$w$ plane (also visible for SDSS~J0100+1818 in Fig.~\ref{fig:cosmo_corner}) is less pronounced, as the redshift of the deflector increases. Only one of the clusters they studied is at a redshift larger than that of SDSS~J0100+1818. This means that our system allows us to exploit still relatively uncommon deflector and source redshift combinations, thus to explore different regions in the $\Omega_{\rm m}$-$w$ plane. These evidences suggest that, together with the comparable accuracy level reached, massive galaxy- and group-scale systems can complement cosmographic studies through strong lensing performed with clusters.

\section{Summary and conclusions}
\label{sec:conclusions}
In this paper, we have exploited new VLT/MUSE data to expand the study of \citet{Bolamperti2023} \citepalias{Bolamperti2023} about the strong lensing system SDSS~J0100+1818. This system consists of a group of galaxies, dominated by a central massive early-type galaxy (BGG), at redshift $z=0.581$, which acts as a strong gravitational deflector for five background sources, all spectroscopically confirmed. The sources are lensed into a total of 18 multiple images, located between 2.25\arcsec\ ($\approx 15$~kpc) and 11.70\arcsec\ ($\approx 77$~kpc) in projection from the BGG. 

We have built, with the software {\tt GLEE}, different strong lensing models of the deflector, consisting in an extended component (whose total mass distribution has been parametrized with either a PIEMD or a SPEMD profile), representing the DM halo in which the group is immersed, and several dPIE components for the member galaxies. We have measured the best-fit parameter values, and their posterior probability distribution, by reproducing the observed position of all the 18 multiple images. 

Taking advantage of the multi-plane nature of the system, with several sources at different redshifts lensed by the same deflector, we have developed some additional models varying also the geometry of the Universe. In fact, by considering the distance ratios between the observer, the lens, and the different sources, and assuming a model of the Universe (flat $\Lambda$CDM or $w$CDM models) the values of $\Omega_{\rm m}$ and $w$ can be inferred.

The main results of this work can be summarized as follows: 
\begin{itemize}
    \item We have confirmed the redshift of the background sources A and B, at $z=1.880$. We have also measured the redshift of source C, at $z = 1.698$. Furthermore, we have discovered two additional background sources, E and F at $z=4.95$, that do not show any HST continuum, but do have clear \lya\ emission lines. \\
    \item We have securely measured the redshifts of 65 objects in the field of view, identifying 19 group members (including the BGG) on the deflector plane, 3 stars, 10 foreground galaxies, 32 background galaxies, and 1 background quasar. \\
     \item We have measured the stellar velocity dispersion profile of the BGG over six bins with $S/N>15$ up to an aperture of 3\arcsec\ radius ($\approx 20$~kpc, approximately twice the value of the effective radius). The profile has a peak of ($386 \pm 6) \, \si{km.s^{-1}}$, and slightly decreases to ($350 \pm 25 )\, \si{km.s^{-1}}$ in the outer regions. \\
    \item We have developed different strong lensing models for the system. We have modeled the projected dimensionless surface mass density of each group member as a dPIE mass distribution, and weighed the total mass of each group galaxy based on its luminosity. We have included an additional mass component, representing all the contributions not associated to the group galaxies. In particular, this main group-scale DM halo has been parametrized with either a PIEMD or a SPEMD total mass distribution. Given the dominance of the BGG, we have also considered models in which the total mass of the BGG was still parametrized by a dPIE profile, but not weighed through the scaling relations adopted for all the other members. The results of these models, summarized in Table~\ref{tab:best-fit}, show that we could reproduce the observed positions of the 18 multiple images with a rms value between 0.21\arcsec and 0.23\arcsec. \\
    \item We have accurately measured a total mass profile for the deflector with an average relative uncertainty of only 2\% between 15 and 77~kpc. In particular, we have measured a total mass of $(1.55 \pm 0.01) \times 10^{13}$~M$_\odot$ ($(2.78 \pm 0.04) \times 10^{13}$~M$_\odot$) within an aperture of $R = 50$~kpc ($R \approx 77$~kpc). The total mass of the extended halo within the same apertures are, respectively, $9.2^{+1.2}_{-3.1} \times 10^{12}$~M$_\odot$ and $1.9^{+0.2}_{-0.7} \times 10^{13}$~M$_\odot$. \\
    \item By adopting the stellar mass-to-light ratio estimated in \citetalias{Bolamperti2023}, we have measured the group cumulative stellar mass profile up to $R \approx 30\arcsec \approx 180$~kpc. We have employed this profile to obtain the stellar over total mass fraction profiles for each model. They have resulted in a value of $45.6^{+8.7}_{-8.3} \%$ at the BGG effective radius, decreasing to $(11.1\pm 1.9) \%$ at $R=50$~kpc, and to $(6.6\pm 1.1) \%$ at $R \approx 77$~kpc. \\
    \item We have developed strong lensing models with the values of $\Omega_{\rm m}$ and $w$ free to vary, in flat $\Lambda$CDM and $w$CDM models for the Universe. The average 1$\sigma$ uncertainty is on the order of 0.1 for $\Omega_{\rm m}$, and of 0.4 for $w$. We have measured $\Omega_{\rm m} = 0.14^{+0.16}_{-0.09}$ (with a 95.4\% upper limit of 0.47) in the flat $\Lambda$CDM case, and $\Omega_{\rm m} = 0.19^{+0.17}_{-0.10}$ (with a 95.4\% upper limit of 0.59) and %$w$ with a $\approx 30\%$ uncertainty, 
    $w = -1.27_{-0.48}^{+0.43}$ (with a 95.4\% upper limit of $-0.53$) in the flat $w$CDM case. \\
    \item We have considered the \mbox{MP-\piemdOm} and \mbox{MP-\piemdOmw} models, by adopting a multi-plane strong lensing modeling approach. These models have included an additional PIEMD mass distribution at $z=1.880$, representing the total mass term associated to sources A and B, and contributing to the total deflection angle. We have found consistent median values for the cosmological parameters $\Omega_{\rm m}$ and $w$, with a similar uncertainty on $w$ and an increased uncertainty on $\Omega_{\rm m}$ by a factor of approximately 1.5. \\
    \item We have checked that the accuracy on the cosmological measurements obtained with SDSS~J0100+1818 is similar to that obtained with the galaxy-scale multi-source-plane strong lensing system known as the ``Jackpot'' \citep{Collett2014, Collett2020, Smith2021}. Given the group nature of SDSS~J0100+1818 and the quite large number of multiple images, we have also compared the accuracy of our estimates with those provided by lens clusters of galaxies. By comparing with the sample presented in \citet{Caminha2022}, we have found that SDSS~J0100+1818 performs as a typical cluster of galaxies, showing that the lower number of multiple images is balanced by the simpler total mass distribution. As an advantage, SDSS~J0100+1818 has a lens redshift that is larger than the average one for the lens clusters, and this reduces the degeneracy between the values of $\Omega_{\rm m}$ and $w$. 
\end{itemize}

The main goals of this study were to make public a new spectroscopic catalog in the SDSS~J0100+1818 field of view, obtained from recent MUSE data, with a particular focus on the deflector group members and on the multiply lensed background sources, and to present the capabilities of this system through a detailed strong lensing analysis to characterize the deflector total mass and infer the values of some cosmological parameters. We plan to enhance this study with already available and future observations. For instance, the kinematics of the BGG and of the brightest group members could be included to refine the strong lensing analysis, new models could be developed considering not just  the point-like position but also the extended structure of the multiple images, and the background sources could be further characterized, by using their morphology and kinematics. Moreover, the extended radial profile shown in Fig.~\ref{fig:sigma_profile} will allow us to develop a joint strong lensing and dynamics model, to better characterize the total mass distribution in the inner regions of the group. Additional new deep rest-frame infrared spectroscopic observations could allow us to measure the redshift of source D, making it the first group-scale system, dominated by a central BGG, with 6 lensed sources on 4 different redshift planes, and possibly discover other distant lensed line emitters, like E and F. Furthermore, deep X-ray observations could confirm the fossil nature of SDSS~J0100+1818, and enable us to put constraints on the extended halo in which the group is immersed. A detailed study of SDSS~J0100+1818 will pave the way to the characterization of tens of such rare galaxy and group scale systems that will be employed for cosmography and that will be discovered with new and upcoming facilities, like Euclid and Rubin-LSST. 

\begin{acknowledgements}
This research is based on observations collected at the European Organisation for Astronomical Research in the Southern Hemisphere under ESO programme 0110.A-0248(A). It is also based on observations collected under ESO programme 091.A-0852(A) and on observations made with the NASA/ESA Hubble Space Telescope obtained from the Space Telescope Science Institute, which is operated by the Association of Universities for Research in Astronomy, Inc., under NASA contract NAS 5–26555. These observations are associated with program GO-15253. We acknowledge the use of \texttt{numpy} \citep{numpy}, \texttt{matplotlib} \citep{matplotlib}, \texttt{astropy} \citep{astropy}, \texttt{pandas} \citep{pandas}, \texttt{corner.py} \citep{corner}, and \texttt{CARTA} \citep{carta_2021}. 
A.B. and A.Z. acknowledge support from the INAF minigrant 1.05.23.04.01 ``Clumps at cosmological distance: revealing their formation, nature, and evolution''.
C.G. and G.G. acknowledge support through grant PRIN-MIUR 2020SKSTHZ. S.H.S.~thanks the Max Planck Society for support through the Max Planck Research Group. L.C. is supported by DFF/Independent Research Fund Denmark, grant-ID 2032–00071.
\end{acknowledgements}

%\begin{appendix}
%\section{Spectrum of the multiple images of the source D}
%\label{app_a}
%\begin{figure*}
%   \centering
%   \includegraphics[width=0.95\textwidth]{figures/specD.jpg}
%      \caption{MUSE 1D spectra of the multiple image D1 and D4, extracted within the purple apertures in Fig.~\ref{fig:regions}. It is smoothed with a boxcar filter. We show zooms in three wavelength ranges where we observe the main absorption and emission lines detected, labeled with their respective interpretation. The purple line represents the observed spectra in units of $10^{-18}$ erg s$^{-1}$ cm$^{-2}$ $\AA^{-1}$, and the gray regions indicate the 1$\sigma$ uncertainties.}
%         \label{fig:specD}
%   \end{figure*}
%\end{appendix}

\bibliographystyle{aa} % style aa.bst
\bibliography{bibliografia} % your references Yourfile.bib

\begin{thebibliography}{98}
\expandafter\ifx\csname natexlab\endcsname\relax\def\natexlab#1{#1}\fi

\bibitem[{{Acebron} {et~al.}(2022){Acebron}, {Grillo}, {Bergamini}, {Caminha}, {Tozzi}, {Mercurio}, {Rosati}, {Brammer}, {Meneghetti}, {Nonino}, \& {Vanzella}}]{Acebron2022}
{Acebron}, A., {Grillo}, C., {Bergamini}, P., {et~al.} 2022, \aap, 668, A142

\bibitem[{{Acebron} {et~al.}(2017){Acebron}, {Jullo}, {Limousin}, {Tilquin}, {Giocoli}, {Jauzac}, {Mahler}, \& {Richard}}]{Acebron2017}
{Acebron}, A., {Jullo}, E., {Limousin}, M., {et~al.} 2017, \mnras, 470, 1809

\bibitem[{{Aguerri} \& {Zarattini}(2021)}]{Aguerri2021}
{Aguerri}, J. A.~L. \& {Zarattini}, S. 2021, Universe, 7, 132

\bibitem[{{Akaike}(1974)}]{Akaike1974}
{Akaike}, H. 1974, IEEE Transactions on Automatic Control, 19, 716

\bibitem[{{Astropy Collaboration} {et~al.}(2022){Astropy Collaboration}, {Price-Whelan}, {Lim}, {Earl}, {Starkman}, {Bradley}, {Shupe}, {Patil}, {Corrales}, {Brasseur}, {N{\"o}the}, {Donath}, {Tollerud}, {Morris}, {Ginsburg}, {Vaher}, {Weaver}, {Tocknell}, {Jamieson}, {van Kerkwijk}, {Robitaille}, {Merry}, {Bachetti}, {G{\"u}nther}, {Aldcroft}, {Alvarado-Montes}, {Archibald}, {B{\'o}di}, {Bapat}, {Barentsen}, {Baz{\'a}n}, {Biswas}, {Boquien}, {Burke}, {Cara}, {Cara}, {Conroy}, {Conseil}, {Craig}, {Cross}, {Cruz}, {D'Eugenio}, {Dencheva}, {Devillepoix}, {Dietrich}, {Eigenbrot}, {Erben}, {Ferreira}, {Foreman-Mackey}, {Fox}, {Freij}, {Garg}, {Geda}, {Glattly}, {Gondhalekar}, {Gordon}, {Grant}, {Greenfield}, {Groener}, {Guest}, {Gurovich}, {Handberg}, {Hart}, {Hatfield-Dodds}, {Homeier}, {Hosseinzadeh}, {Jenness}, {Jones}, {Joseph}, {Kalmbach}, {Karamehmetoglu}, {Ka{\l}uszy{\'n}ski}, {Kelley}, {Kern}, {Kerzendorf}, {Koch}, {Kulumani}, {Lee}, {Ly}, {Ma}, {MacBride}, {Maljaars}, {Muna}, {Murphy}, {Norman},
  {O'Steen}, {Oman}, {Pacifici}, {Pascual}, {Pascual-Granado}, {Patil}, {Perren}, {Pickering}, {Rastogi}, {Roulston}, {Ryan}, {Rykoff}, {Sabater}, {Sakurikar}, {Salgado}, {Sanghi}, {Saunders}, {Savchenko}, {Schwardt}, {Seifert-Eckert}, {Shih}, {Jain}, {Shukla}, {Sick}, {Simpson}, {Singanamalla}, {Singer}, {Singhal}, {Sinha}, {Sip{\H{o}}cz}, {Spitler}, {Stansby}, {Streicher}, {{\v{S}}umak}, {Swinbank}, {Taranu}, {Tewary}, {Tremblay}, {de Val-Borro}, {Van Kooten}, {Vasovi{\'c}}, {Verma}, {de Miranda Cardoso}, {Williams}, {Wilson}, {Winkel}, {Wood-Vasey}, {Xue}, {Yoachim}, {Zhang}, {Zonca}, \& {Astropy Project Contributors}}]{astropy}
{Astropy Collaboration}, {Price-Whelan}, A.~M., {Lim}, P.~L., {et~al.} 2022, \apj, 935, 167

\bibitem[{{Bacon} {et~al.}(2023){Bacon}, {Brinchmann}, {Conseil}, {Maseda}, {Nanayakkara}, {Wendt}, {Bacher}, {Mary}, {Weilbacher}, {Krajnovi{\'c}}, {Boogaard}, {Bouch{\'e}}, {Contini}, {Epinat}, {Feltre}, {Guo}, {Herenz}, {Kollatschny}, {Kusakabe}, {Leclercq}, {Michel-Dansac}, {Pello}, {Richard}, {Roth}, {Salvignol}, {Schaye}, {Steinmetz}, {Tresse}, {Urrutia}, {Verhamme}, {Vitte}, {Wisotzki}, \& {Zoutendijk}}]{Bacon2023}
{Bacon}, R., {Brinchmann}, J., {Conseil}, S., {et~al.} 2023, \aap, 670, A4

\bibitem[{{Balestra} {et~al.}(2016){Balestra}, {Mercurio}, {Sartoris}, {Girardi}, {Grillo}, {Nonino}, {Rosati}, {Biviano}, {Ettori}, {Forman}, {Jones}, {Koekemoer}, {Medezinski}, {Merten}, {Ogrean}, {Tozzi}, {Umetsu}, {Vanzella}, {van Weeren}, {Zitrin}, {Annunziatella}, {Caminha}, {Broadhurst}, {Coe}, {Donahue}, {Fritz}, {Frye}, {Kelson}, {Lombardi}, {Maier}, {Meneghetti}, {Monna}, {Postman}, {Scodeggio}, {Seitz}, \& {Ziegler}}]{Balestra2016}
{Balestra}, I., {Mercurio}, A., {Sartoris}, B., {et~al.} 2016, \apjs, 224, 33

\bibitem[{{Ballard} {et~al.}(2024){Ballard}, {Enzi}, {Collett}, {Turner}, \& {Smith}}]{Ballard2024}
{Ballard}, D.~J., {Enzi}, W. J.~R., {Collett}, T.~E., {Turner}, H.~C., \& {Smith}, R.~J. 2024, \mnras, 528, 7564

\bibitem[{{Barkana}(1998)}]{Barkana1998}
{Barkana}, R. 1998, \apj, 502, 531

\bibitem[{{Bartelmann}(2010)}]{Bartelmann2010}
{Bartelmann}, M. 2010, Classical and Quantum Gravity, 27, 233001

\bibitem[{{Belokurov} {et~al.}(2009){Belokurov}, {Evans}, {Hewett}, {Moiseev}, {McMahon}, {Sanchez}, \& {King}}]{Belokurov2009}
{Belokurov}, V., {Evans}, N.~W., {Hewett}, P.~C., {et~al.} 2009, \mnras, 392, 104

\bibitem[{{Bender} {et~al.}(1992){Bender}, {Burstein}, \& {Faber}}]{Bender1992}
{Bender}, R., {Burstein}, D., \& {Faber}, S.~M. 1992, \apj, 399, 462

\bibitem[{{Bergamini} {et~al.}(2023{\natexlab{a}}){Bergamini}, {Acebron}, {Grillo}, {Rosati}, {Caminha}, {Mercurio}, {Vanzella}, {Mason}, {Treu}, {Angora}, {Brammer}, {Meneghetti}, {Nonino}, {Boyett}, {Brada{\v{c}}}, {Castellano}, {Fontana}, {Morishita}, {Paris}, {Prieto-Lyon}, {Roberts-Borsani}, {Roy}, {Santini}, {Vulcani}, {Wang}, \& {Yang}}]{Bergamini2023b}
{Bergamini}, P., {Acebron}, A., {Grillo}, C., {et~al.} 2023{\natexlab{a}}, \apj, 952, 84

\bibitem[{{Bergamini} {et~al.}(2023{\natexlab{b}}){Bergamini}, {Grillo}, {Rosati}, {Vanzella}, {Me{\v{s}}tri{\'c}}, {Mercurio}, {Acebron}, {Caminha}, {Granata}, {Meneghetti}, {Angora}, \& {Nonino}}]{Bergamini2023}
{Bergamini}, P., {Grillo}, C., {Rosati}, P., {et~al.} 2023{\natexlab{b}}, \aap, 674, A79

\bibitem[{{Bernardi} {et~al.}(2003){Bernardi}, {Sheth}, {Annis}, {Burles}, {Eisenstein}, {Finkbeiner}, {Hogg}, {Lupton}, {Schlegel}, {SubbaRao}, {Bahcall}, {Blakeslee}, {Brinkmann}, {Castander}, {Connolly}, {Csabai}, {Doi}, {Fukugita}, {Frieman}, {Heckman}, {Hennessy}, {Ivezi{\'c}}, {Knapp}, {Lamb}, {McKay}, {Munn}, {Nichol}, {Okamura}, {Schneider}, {Thakar}, \& {York}}]{Bernardi2003}
{Bernardi}, M., {Sheth}, R.~K., {Annis}, J., {et~al.} 2003, \aj, 125, 1866

\bibitem[{{Bertin} \& {Arnouts}(1996)}]{Bertin1996}
{Bertin}, E. \& {Arnouts}, S. 1996, \aaps, 117, 393

\bibitem[{{Birrer} {et~al.}(2019){Birrer}, {Treu}, {Rusu}, {Bonvin}, {Fassnacht}, {Chan}, {Agnello}, {Shajib}, {Chen}, {Auger}, {Courbin}, {Hilbert}, {Sluse}, {Suyu}, {Wong}, {Marshall}, {Lemaux}, \& {Meylan}}]{Birrer2019}
{Birrer}, S., {Treu}, T., {Rusu}, C.~E., {et~al.} 2019, \mnras, 484, 4726

\bibitem[{{Blandford} \& {Narayan}(1986)}]{Blandford1986}
{Blandford}, R. \& {Narayan}, R. 1986, \apj, 310, 568

\bibitem[{{Bolamperti} {et~al.}(2023){Bolamperti}, {Grillo}, {Ca{\~n}ameras}, {Suyu}, \& {Christensen}}]{Bolamperti2023}
{Bolamperti}, A., {Grillo}, C., {Ca{\~n}ameras}, R., {Suyu}, S.~H., \& {Christensen}, L. 2023, \aap, 671, A60

\bibitem[{{Caminha} {et~al.}(2016){Caminha}, {Grillo}, {Rosati}, {Balestra}, {Karman}, {Lombardi}, {Mercurio}, {Nonino}, {Tozzi}, {Zitrin}, {Biviano}, {Girardi}, {Koekemoer}, {Melchior}, {Meneghetti}, {Munari}, {Suyu}, {Umetsu}, {Annunziatella}, {Borgani}, {Broadhurst}, {Caputi}, {Coe}, {Delgado-Correal}, {Ettori}, {Fritz}, {Frye}, {Gobat}, {Maier}, {Monna}, {Postman}, {Sartoris}, {Seitz}, {Vanzella}, \& {Ziegler}}]{Caminha2016}
{Caminha}, G.~B., {Grillo}, C., {Rosati}, P., {et~al.} 2016, \aap, 587, A80

\bibitem[{{Caminha} {et~al.}(2019){Caminha}, {Rosati}, {Grillo}, {Rosani}, {Caputi}, {Meneghetti}, {Mercurio}, {Balestra}, {Bergamini}, {Biviano}, {Nonino}, {Umetsu}, {Vanzella}, {Annunziatella}, {Broadhurst}, {Delgado-Correal}, {Demarco}, {Koekemoer}, {Lombardi}, {Maier}, {Verdugo}, \& {Zitrin}}]{Caminha2019}
{Caminha}, G.~B., {Rosati}, P., {Grillo}, C., {et~al.} 2019, \aap, 632, A36

\bibitem[{{Caminha} {et~al.}(2022){Caminha}, {Suyu}, {Grillo}, \& {Rosati}}]{Caminha2022}
{Caminha}, G.~B., {Suyu}, S.~H., {Grillo}, C., \& {Rosati}, P. 2022, \aap, 657, A83

\bibitem[{{Cao} {et~al.}(2012){Cao}, {Pan}, {Biesiada}, {Godlowski}, \& {Zhu}}]{Cao2012}
{Cao}, S., {Pan}, Y., {Biesiada}, M., {Godlowski}, W., \& {Zhu}, Z.-H. 2012, \jcap, 2012, 016

\bibitem[{{Cappellari}(2017)}]{Cappellari17}
{Cappellari}, M. 2017, \mnras, 466, 798

\bibitem[{{Cappellari}(2023)}]{Cappellari23}
{Cappellari}, M. 2023, \mnras, 526, 3273

\bibitem[{{Cappellari} \& {Emsellem}(2004)}]{Cappellari04}
{Cappellari}, M. \& {Emsellem}, E. 2004, \pasp, 116, 138

\bibitem[{Cavanaugh(1997)}]{CAVANAUGH1997201}
Cavanaugh, J.~E. 1997, Statistics \& Probability Letters, 33, 201

\bibitem[{{Chiriv{\`\i}} {et~al.}(2018){Chiriv{\`\i}}, {Suyu}, {Grillo}, {Halkola}, {Balestra}, {Caminha}, {Mercurio}, \& {Rosati}}]{Chirivi2018}
{Chiriv{\`\i}}, G., {Suyu}, S.~H., {Grillo}, C., {et~al.} 2018, \aap, 614, A8

\bibitem[{{Ciotti} {et~al.}(1996){Ciotti}, {Lanzoni}, \& {Renzini}}]{Ciotti1996}
{Ciotti}, L., {Lanzoni}, B., \& {Renzini}, A. 1996, \mnras, 282, 1

\bibitem[{{Collett} \& {Auger}(2014)}]{Collett2014}
{Collett}, T.~E. \& {Auger}, M.~W. 2014, \mnras, 443, 969

\bibitem[{{Collett} {et~al.}(2012){Collett}, {Auger}, {Belokurov}, {Marshall}, \& {Hall}}]{Collett2012}
{Collett}, T.~E., {Auger}, M.~W., {Belokurov}, V., {Marshall}, P.~J., \& {Hall}, A.~C. 2012, \mnras, 424, 2864

\bibitem[{{Collett} \& {Smith}(2020)}]{Collett2020}
{Collett}, T.~E. \& {Smith}, R.~J. 2020, \mnras, 497, 1654

\bibitem[{Comrie {et~al.}(2021)Comrie, Wang, Hsu, Moraghan, Harris, Pang, Pińska, Chiang, Chang, Hwang, Jan, Lin, \& Simmonds}]{carta_2021}
Comrie, A., Wang, K.-S., Hsu, S.-C., {et~al.} 2021, {CARTA: The Cube Analysis and Rendering Tool for Astronomy}

\bibitem[{{Dariush} {et~al.}(2010){Dariush}, {Raychaudhury}, {Ponman}, {Khosroshahi}, {Benson}, {Bower}, \& {Pearce}}]{Dariush2010}
{Dariush}, A.~A., {Raychaudhury}, S., {Ponman}, T.~J., {et~al.} 2010, \mnras, 405, 1873

\bibitem[{{Eisenstein} {et~al.}(2005){Eisenstein}, {Zehavi}, {Hogg}, {Scoccimarro}, {Blanton}, {Nichol}, {Scranton}, {Seo}, {Tegmark}, {Zheng}, {Anderson}, {Annis}, {Bahcall}, {Brinkmann}, {Burles}, {Castander}, {Connolly}, {Csabai}, {Doi}, {Fukugita}, {Frieman}, {Glazebrook}, {Gunn}, {Hendry}, {Hennessy}, {Ivezi{\'c}}, {Kent}, {Knapp}, {Lin}, {Loh}, {Lupton}, {Margon}, {McKay}, {Meiksin}, {Munn}, {Pope}, {Richmond}, {Schlegel}, {Schneider}, {Shimasaku}, {Stoughton}, {Strauss}, {SubbaRao}, {Szalay}, {Szapudi}, {Tucker}, {Yanny}, \& {York}}]{Eisenstein2005}
{Eisenstein}, D.~J., {Zehavi}, I., {Hogg}, D.~W., {et~al.} 2005, \apj, 633, 560

\bibitem[{{El{\'\i}asd{\'o}ttir} {et~al.}(2007){El{\'\i}asd{\'o}ttir}, {Limousin}, {Richard}, {Hjorth}, {Kneib}, {Natarajan}, {Pedersen}, {Jullo}, \& {Paraficz}}]{Eliasdottir2007}
{El{\'\i}asd{\'o}ttir}, {\'A}., {Limousin}, M., {Richard}, J., {et~al.} 2007, arXiv e-prints, arXiv:0710.5636

\bibitem[{{ESO CPL Development Team}(2015)}]{esorex}
{ESO CPL Development Team}. 2015, {EsoRex: ESO Recipe Execution Tool}, Astrophysics Source Code Library, record ascl:1504.003

\bibitem[{{Faber} {et~al.}(1987){Faber}, {Dressler}, {Davies}, {Burstein}, {Lynden Bell}, {Terlevich}, \& {Wegner}}]{Faber1987}
{Faber}, S.~M., {Dressler}, A., {Davies}, R.~L., {et~al.} 1987, in Nearly Normal Galaxies. From the Planck Time to the Present, ed. S.~M. {Faber}, 175

\bibitem[{Foreman-Mackey(2016)}]{corner}
Foreman-Mackey, D. 2016, The Journal of Open Source Software, 1, 24

\bibitem[{{Gavazzi} {et~al.}(2008){Gavazzi}, {Treu}, {Koopmans}, {Bolton}, {Moustakas}, {Burles}, \& {Marshall}}]{Gavazzi2008}
{Gavazzi}, R., {Treu}, T., {Koopmans}, L. V.~E., {et~al.} 2008, \apj, 677, 1046

\bibitem[{{Gnedin} {et~al.}(2004){Gnedin}, {Kravtsov}, {Klypin}, \& {Nagai}}]{Gnedin2004}
{Gnedin}, O.~Y., {Kravtsov}, A.~V., {Klypin}, A.~A., \& {Nagai}, D. 2004, \apj, 616, 16

\bibitem[{{Gonneau} {et~al.}(2020){Gonneau}, {Lyubenova}, {Lan{\c{c}}on}, {Trager}, {Peletier}, {Arentsen}, {Chen}, {Coelho}, {Dries}, {Falc{\'o}n-Barroso}, {Prugniel}, {S{\'a}nchez-Bl{\'a}zquez}, {Vazdekis}, \& {Verro}}]{gonneau20}
{Gonneau}, A., {Lyubenova}, M., {Lan{\c{c}}on}, A., {et~al.} 2020, \aap, 634, A133

\bibitem[{{Granata} {et~al.}(2023){Granata}, {Bergamini}, {Grillo}, {Meneghetti}, {Mercurio}, {Me{\v{s}}tri{\'c}}, {Ragagnin}, {Rosati}, {Caminha}, {Tortorelli}, \& {Vanzella}}]{Granata2023}
{Granata}, G., {Bergamini}, P., {Grillo}, C., {et~al.} 2023, \aap, 679, A124

\bibitem[{{Grillo} {et~al.}(2013){Grillo}, {Christensen}, {Gallazzi}, \& {Rasmussen}}]{Grillo2013}
{Grillo}, C., {Christensen}, L., {Gallazzi}, A., \& {Rasmussen}, J. 2013, \mnras, 433, 2604

\bibitem[{{Grillo} \& {Gobat}(2010)}]{GrilloGobat2010}
{Grillo}, C. \& {Gobat}, R. 2010, \mnras, 402, L67

\bibitem[{{Grillo} {et~al.}(2008){Grillo}, {Lombardi}, \& {Bertin}}]{Grillo2008}
{Grillo}, C., {Lombardi}, M., \& {Bertin}, G. 2008, \aap, 477, 397

\bibitem[{{Grillo} {et~al.}(2024){Grillo}, {Pagano}, {Rosati}, \& {Suyu}}]{Grillo2024}
{Grillo}, C., {Pagano}, L., {Rosati}, P., \& {Suyu}, S.~H. 2024, \aap, 684, L23

\bibitem[{{Grillo} {et~al.}(2018){Grillo}, {Rosati}, {Suyu}, {Balestra}, {Caminha}, {Halkola}, {Kelly}, {Lombardi}, {Mercurio}, {Rodney}, \& {Treu}}]{Grillo2018}
{Grillo}, C., {Rosati}, P., {Suyu}, S.~H., {et~al.} 2018, \apj, 860, 94

\bibitem[{{Grillo} {et~al.}(2020){Grillo}, {Rosati}, {Suyu}, {Caminha}, {Mercurio}, \& {Halkola}}]{Grillo2020}
{Grillo}, C., {Rosati}, P., {Suyu}, S.~H., {et~al.} 2020, \apj, 898, 87

\bibitem[{{Grillo} {et~al.}(2015){Grillo}, {Suyu}, {Rosati}, {Mercurio}, {Balestra}, {Munari}, {Nonino}, {Caminha}, {Lombardi}, {De Lucia}, {Borgani}, {Gobat}, {Biviano}, {Girardi}, {Umetsu}, {Coe}, {Koekemoer}, {Postman}, {Zitrin}, {Halkola}, {Broadhurst}, {Sartoris}, {Presotto}, {Annunziatella}, {Maier}, {Fritz}, {Vanzella}, \& {Frye}}]{Grillo2015}
{Grillo}, C., {Suyu}, S.~H., {Rosati}, P., {et~al.} 2015, \apj, 800, 38

\bibitem[{Harris {et~al.}(2020)Harris, Millman, van~der Walt, Gommers, Virtanen, Cournapeau, Wieser, Taylor, Berg, Smith, Kern, Picus, Hoyer, van Kerkwijk, Brett, Haldane, del R{\'{i}}o, Wiebe, Peterson, G{\'{e}}rard-Marchant, Sheppard, Reddy, Weckesser, Abbasi, Gohlke, \& Oliphant}]{numpy}
Harris, C.~R., Millman, K.~J., van~der Walt, S.~J., {et~al.} 2020, Nature, 585, 357

\bibitem[{{Hinton} {et~al.}(2016){Hinton}, {Davis}, {Lidman}, {Glazebrook}, \& {Lewis}}]{Hinton2016Marz}
{Hinton}, S.~R., {Davis}, T.~M., {Lidman}, C., {Glazebrook}, K., \& {Lewis}, G.~F. 2016, Astronomy and Computing, 15, 61

\bibitem[{Hunter(2007)}]{matplotlib}
Hunter, J.~D. 2007, Computing in Science \& Engineering, 9, 90

\bibitem[{{Johnson} {et~al.}(2018{\natexlab{a}}){Johnson}, {Irwin}, {White}, {Wong}, \& {Dupke}}]{Johnson2018b}
{Johnson}, L.~E., {Irwin}, J.~A., {White}, Raymond~E., I., {Wong}, K.-W., \& {Dupke}, R.~A. 2018{\natexlab{a}}, \apj, 869, 170

\bibitem[{{Johnson} {et~al.}(2018{\natexlab{b}}){Johnson}, {Irwin}, {White}, {Wong}, {Maksym}, {Dupke}, {Miller}, \& {Carrasco}}]{Johnson2018}
{Johnson}, L.~E., {Irwin}, J.~A., {White}, Raymond~E., I., {et~al.} 2018{\natexlab{b}}, \apj, 856, 131

\bibitem[{{Jones} {et~al.}(2003){Jones}, {Ponman}, {Horton}, {Babul}, {Ebeling}, \& {Burke}}]{Jones2003}
{Jones}, L.~R., {Ponman}, T.~J., {Horton}, A., {et~al.} 2003, \mnras, 343, 627

\bibitem[{{Jullo} {et~al.}(2010){Jullo}, {Natarajan}, {Kneib}, {D'Aloisio}, {Limousin}, {Richard}, \& {Schimd}}]{Jullo2010}
{Jullo}, E., {Natarajan}, P., {Kneib}, J.~P., {et~al.} 2010, Science, 329, 924

\bibitem[{{Khosroshahi} {et~al.}(2006){Khosroshahi}, {Maughan}, {Ponman}, \& {Jones}}]{Khosroshahi2006}
{Khosroshahi}, H.~G., {Maughan}, B.~J., {Ponman}, T.~J., \& {Jones}, L.~R. 2006, \mnras, 369, 1211

\bibitem[{{Kneib} {et~al.}(1996){Kneib}, {Ellis}, {Smail}, {Couch}, \& {Sharples}}]{Kneib1996}
{Kneib}, J.~P., {Ellis}, R.~S., {Smail}, I., {Couch}, W.~J., \& {Sharples}, R.~M. 1996, \apj, 471, 643

\bibitem[{{Komatsu} {et~al.}(2014){Komatsu}, {Bennett}, {Barnes}, {Bean}, {Bennett}, {Dor{\'e}}, {Dunkley}, {Gold}, {Greason}, {Halpern}, {Hill}, {Hinshaw}, {Jarosik}, {Kogut}, {Komatsu}, {Larson}, {Limon}, {Meyer}, {Nolta}, {Odegard}, {Page}, {Peiris}, {Smith}, {Spergel}, {Tucker}, {Verde}, {Weiland}, {Wollack}, \& {Wright}}]{Komatsu2014}
{Komatsu}, E., {Bennett}, C.~L., {Barnes}, C., {et~al.} 2014, Progress of Theoretical and Experimental Physics, 2014, 06B102

\bibitem[{{Li} {et~al.}(2024){Li}, {Collett}, {Krawczyk}, \& {Enzi}}]{Li2024}
{Li}, T., {Collett}, T.~E., {Krawczyk}, C.~M., \& {Enzi}, W. 2024, \mnras, 527, 5311

\bibitem[{{Mahler} {et~al.}(2018){Mahler}, {Richard}, {Cl{\'e}ment}, {Lagattuta}, {Schmidt}, {Patr{\'\i}cio}, {Soucail}, {Bacon}, {Pello}, {Bouwens}, {Maseda}, {Martinez}, {Carollo}, {Inami}, {Leclercq}, \& {Wisotzki}}]{Mahler2018}
{Mahler}, G., {Richard}, J., {Cl{\'e}ment}, B., {et~al.} 2018, \mnras, 473, 663

\bibitem[{{Martizzi} {et~al.}(2012){Martizzi}, {Teyssier}, {Moore}, \& {Wentz}}]{Martizzi2012}
{Martizzi}, D., {Teyssier}, R., {Moore}, B., \& {Wentz}, T. 2012, \mnras, 422, 3081

\bibitem[{{Meneghetti} {et~al.}(2020){Meneghetti}, {Davoli}, {Bergamini}, {Rosati}, {Natarajan}, {Giocoli}, {Caminha}, {Metcalf}, {Rasia}, {Borgani}, {Calura}, {Grillo}, {Mercurio}, \& {Vanzella}}]{Meneghetti2020}
{Meneghetti}, M., {Davoli}, G., {Bergamini}, P., {et~al.} 2020, Science, 369, 1347

\bibitem[{{Moresco} {et~al.}(2022){Moresco}, {Amati}, {Amendola}, {Birrer}, {Blakeslee}, {Cantiello}, {Cimatti}, {Darling}, {Della Valle}, {Fishbach}, {Grillo}, {Hamaus}, {Holz}, {Izzo}, {Jimenez}, {Lusso}, {Meneghetti}, {Piedipalumbo}, {Pisani}, {Pourtsidou}, {Pozzetti}, {Quartin}, {Risaliti}, {Rosati}, \& {Verde}}]{Moresco2022}
{Moresco}, M., {Amati}, L., {Amendola}, L., {et~al.} 2022, Living Reviews in Relativity, 25, 6

\bibitem[{{Motta} {et~al.}(2021){Motta}, {Garc{\'\i}a-Aspeitia}, {Hern{\'a}ndez-Almada}, {Maga{\~n}a}, \& {Verdugo}}]{Motta2021}
{Motta}, V., {Garc{\'\i}a-Aspeitia}, M.~A., {Hern{\'a}ndez-Almada}, A., {Maga{\~n}a}, J., \& {Verdugo}, T. 2021, Universe, 7, 163

\bibitem[{{Newman} {et~al.}(2013{\natexlab{a}}){Newman}, {Treu}, {Ellis}, \& {Sand}}]{Newman2013b}
{Newman}, A.~B., {Treu}, T., {Ellis}, R.~S., \& {Sand}, D.~J. 2013{\natexlab{a}}, \apj, 765, 25

\bibitem[{{Newman} {et~al.}(2013{\natexlab{b}}){Newman}, {Treu}, {Ellis}, {Sand}, {Nipoti}, {Richard}, \& {Jullo}}]{Newman2013a}
{Newman}, A.~B., {Treu}, T., {Ellis}, R.~S., {et~al.} 2013{\natexlab{b}}, \apj, 765, 24

\bibitem[{{Oke}(1974)}]{Oke1974}
{Oke}, J.~B. 1974, \apjs, 27, 21

\bibitem[{pandas~development team(2023)}]{pandas}
pandas~development team, T. 2023, pandas-dev/pandas: Pandas

\bibitem[{{Perlmutter} {et~al.}(1999){Perlmutter}, {Aldering}, {Goldhaber}, {Knop}, {Nugent}, {Castro}, {Deustua}, {Fabbro}, {Goobar}, {Groom}, {Hook}, {Kim}, {Kim}, {Lee}, {Nunes}, {Pain}, {Pennypacker}, {Quimby}, {Lidman}, {Ellis}, {Irwin}, {McMahon}, {Ruiz-Lapuente}, {Walton}, {Schaefer}, {Boyle}, {Filippenko}, {Matheson}, {Fruchter}, {Panagia}, {Newberg}, {Couch}, \& {Project}}]{Perlmutter1999}
{Perlmutter}, S., {Aldering}, G., {Goldhaber}, G., {et~al.} 1999, \apj, 517, 565

\bibitem[{{Planck Collaboration} {et~al.}(2014){Planck Collaboration}, {Ade}, {Aghanim}, {Armitage-Caplan}, {Arnaud}, {Ashdown}, {Atrio-Barandela}, {Aumont}, {Baccigalupi}, {Banday}, {Barreiro}, {Bartlett}, {Battaner}, {Benabed}, {Beno{\^\i}t}, {Benoit-L{\'e}vy}, {Bernard}, {Bersanelli}, {Bielewicz}, {Bobin}, {Bock}, {Bonaldi}, {Bond}, {Borrill}, {Bouchet}, {Bridges}, {Bucher}, {Burigana}, {Butler}, {Calabrese}, {Cappellini}, {Cardoso}, {Catalano}, {Challinor}, {Chamballu}, {Chary}, {Chen}, {Chiang}, {Chiang}, {Christensen}, {Church}, {Clements}, {Colombi}, {Colombo}, {Couchot}, {Coulais}, {Crill}, {Curto}, {Cuttaia}, {Danese}, {Davies}, {Davis}, {de Bernardis}, {de Rosa}, {de Zotti}, {Delabrouille}, {Delouis}, {D{\'e}sert}, {Dickinson}, {Diego}, {Dolag}, {Dole}, {Donzelli}, {Dor{\'e}}, {Douspis}, {Dunkley}, {Dupac}, {Efstathiou}, {Elsner}, {En{\ss}lin}, {Eriksen}, {Finelli}, {Forni}, {Frailis}, {Fraisse}, {Franceschi}, {Gaier}, {Galeotta}, {Galli}, {Ganga}, {Giard}, {Giardino}, {Giraud-H{\'e}raud},
  {Gjerl{\o}w}, {Gonz{\'a}lez-Nuevo}, {G{\'o}rski}, {Gratton}, {Gregorio}, {Gruppuso}, {Gudmundsson}, {Haissinski}, {Hamann}, {Hansen}, {Hanson}, {Harrison}, {Henrot-Versill{\'e}}, {Hern{\'a}ndez-Monteagudo}, {Herranz}, {Hildebrandt}, {Hivon}, {Hobson}, {Holmes}, {Hornstrup}, {Hou}, {Hovest}, {Huffenberger}, {Jaffe}, {Jaffe}, {Jewell}, {Jones}, {Juvela}, {Keih{\"a}nen}, {Keskitalo}, {Kisner}, {Kneissl}, {Knoche}, {Knox}, {Kunz}, {Kurki-Suonio}, {Lagache}, {L{\"a}hteenm{\"a}ki}, {Lamarre}, {Lasenby}, {Lattanzi}, {Laureijs}, {Lawrence}, {Leach}, {Leahy}, {Leonardi}, {Le{\'o}n-Tavares}, {Lesgourgues}, {Lewis}, {Liguori}, {Lilje}, {Linden-V{\o}rnle}, {L{\'o}pez-Caniego}, {Lubin}, {Mac{\'\i}as-P{\'e}rez}, {Maffei}, {Maino}, {Mandolesi}, {Maris}, {Marshall}, {Martin}, {Mart{\'\i}nez-Gonz{\'a}lez}, {Masi}, {Massardi}, {Matarrese}, {Matthai}, {Mazzotta}, {Meinhold}, {Melchiorri}, {Melin}, {Mendes}, {Menegoni}, {Mennella}, {Migliaccio}, {Millea}, {Mitra}, {Miville-Desch{\^e}nes}, {Moneti}, {Montier}, {Morgante},
  {Mortlock}, {Moss}, {Munshi}, {Murphy}, {Naselsky}, {Nati}, {Natoli}, {Netterfield}, {N{\o}rgaard-Nielsen}, {Noviello}, {Novikov}, {Novikov}, {O'Dwyer}, {Osborne}, {Oxborrow}, {Paci}, {Pagano}, {Pajot}, {Paladini}, {Paoletti}, {Partridge}, {Pasian}, {Patanchon}, {Pearson}, {Pearson}, {Peiris}, {Perdereau}, {Perotto}, {Perrotta}, {Pettorino}, {Piacentini}, {Piat}, {Pierpaoli}, {Pietrobon}, {Plaszczynski}, {Platania}, {Pointecouteau}, {Polenta}, {Ponthieu}, {Popa}, {Poutanen}, {Pratt}, {Pr{\'e}zeau}, {Prunet}, {Puget}, {Rachen}, {Reach}, {Rebolo}, {Reinecke}, {Remazeilles}, {Renault}, {Ricciardi}, {Riller}, {Ristorcelli}, {Rocha}, {Rosset}, {Roudier}, {Rowan-Robinson}, {Rubi{\~n}o-Mart{\'\i}n}, {Rusholme}, {Sandri}, {Santos}, {Savelainen}, {Savini}, {Scott}, {Seiffert}, {Shellard}, {Spencer}, {Starck}, {Stolyarov}, {Stompor}, {Sudiwala}, {Sunyaev}, {Sureau}, {Sutton}, {Suur-Uski}, {Sygnet}, {Tauber}, {Tavagnacco}, {Terenzi}, {Toffolatti}, {Tomasi}, {Tristram}, {Tucci}, {Tuovinen}, {T{\"u}rler}, {Umana},
  {Valenziano}, {Valiviita}, {Van Tent}, {Vielva}, {Villa}, {Vittorio}, {Wade}, {Wandelt}, {Wehus}, {White}, {White}, {Wilkinson}, {Yvon}, {Zacchei}, \& {Zonca}}]{Planck2014}
{Planck Collaboration}, {Ade}, P.~A.~R., {Aghanim}, N., {et~al.} 2014, \aap, 571, A16

\bibitem[{{Ponman} \& {Bertram}(1993)}]{Ponman1993}
{Ponman}, T.~J. \& {Bertram}, D. 1993, \nat, 363, 51

\bibitem[{{Refsdal}(1964)}]{Refsdal1964}
{Refsdal}, S. 1964, \mnras, 128, 307

\bibitem[{{Riess} {et~al.}(1998){Riess}, {Filippenko}, {Challis}, {Clocchiatti}, {Diercks}, {Garnavich}, {Gilliland}, {Hogan}, {Jha}, {Kirshner}, {Leibundgut}, {Phillips}, {Reiss}, {Schmidt}, {Schommer}, {Smith}, {Spyromilio}, {Stubbs}, {Suntzeff}, \& {Tonry}}]{Riess1998}
{Riess}, A.~G., {Filippenko}, A.~V., {Challis}, P., {et~al.} 1998, \aj, 116, 1009

\bibitem[{{Rusu} {et~al.}(2020){Rusu}, {Wong}, {Bonvin}, {Sluse}, {Suyu}, {Fassnacht}, {Chan}, {Hilbert}, {Auger}, {Sonnenfeld}, {Birrer}, {Courbin}, {Treu}, {Chen}, {Halkola}, {Koopmans}, {Marshall}, \& {Shajib}}]{Rusu2020}
{Rusu}, C.~E., {Wong}, K.~C., {Bonvin}, V., {et~al.} 2020, \mnras, 498, 1440

\bibitem[{{Schneider}(2014)}]{Schneider2014}
{Schneider}, P. 2014, \aap, 568, L2

\bibitem[{{Schneider} {et~al.}(1992){Schneider}, {Ehlers}, \& {Falco}}]{Schneider1992}
{Schneider}, P., {Ehlers}, J., \& {Falco}, E.~E. 1992, {Gravitational Lenses}

\bibitem[{{Schuldt} {et~al.}(2019){Schuldt}, {Chiriv{\`\i}}, {Suyu}, {Y{\i}ld{\i}r{\i}m}, {Sonnenfeld}, {Halkola}, \& {Lewis}}]{Schuldt2019}
{Schuldt}, S., {Chiriv{\`\i}}, G., {Suyu}, S.~H., {et~al.} 2019, \aap, 631, A40

\bibitem[{{Schuldt} {et~al.}(2024){Schuldt}, {Grillo}, {Caminha}, {Mercurio}, {Rosati}, {Morishita}, {Stiavelli}, {Suyu}, {Bergamini}, {Brescia}, {Calura}, \& {Meneghetti}}]{Schuldt2024}
{Schuldt}, S., {Grillo}, C., {Caminha}, G.~B., {et~al.} 2024, \aap, 689, A42

\bibitem[{{Schwarz}(1978)}]{Schwarz1978}
{Schwarz}, G. 1978, Annals of Statistics, 6, 461

\bibitem[{{Shajib} {et~al.}(2023){Shajib}, {Mozumdar}, {Chen}, {Treu}, {Cappellari}, {Knabel}, {Suyu}, {Bennert}, {Frieman}, {Sluse}, {Birrer}, {Courbin}, {Fassnacht}, {Villafa{\~n}a}, \& {Williams}}]{Shajib2023}
{Shajib}, A.~J., {Mozumdar}, P., {Chen}, G. C.~F., {et~al.} 2023, \aap, 673, A9

\bibitem[{{Smith} \& {Collett}(2021)}]{Smith2021}
{Smith}, R.~J. \& {Collett}, T.~E. 2021, \mnras, 505, 2136

\bibitem[{{Soto} {et~al.}(2016){Soto}, {Lilly}, {Bacon}, {Richard}, \& {Conseil}}]{Soto2016}
{Soto}, K.~T., {Lilly}, S.~J., {Bacon}, R., {Richard}, J., \& {Conseil}, S. 2016, \mnras, 458, 3210

\bibitem[{{Stark} {et~al.}(2013){Stark}, {Auger}, {Belokurov}, {Jones}, {Robertson}, {Ellis}, {Sand}, {Moiseev}, {Eagle}, \& {Myers}}]{Stark2013}
{Stark}, D.~P., {Auger}, M., {Belokurov}, V., {et~al.} 2013, \mnras, 436, 1040

\bibitem[{{Suyu} {et~al.}(2017){Suyu}, {Bonvin}, {Courbin}, {Fassnacht}, {Rusu}, {Sluse}, {Treu}, {Wong}, {Auger}, {Ding}, {Hilbert}, {Marshall}, {Rumbaugh}, {Sonnenfeld}, {Tewes}, {Tihhonova}, {Agnello}, {Blandford}, {Chen}, {Collett}, {Koopmans}, {Liao}, {Meylan}, \& {Spiniello}}]{Suyu2017}
{Suyu}, S.~H., {Bonvin}, V., {Courbin}, F., {et~al.} 2017, \mnras, 468, 2590

\bibitem[{{Suyu} \& {Halkola}(2010)}]{Suyu2010}
{Suyu}, S.~H. \& {Halkola}, A. 2010, \aap, 524, A94

\bibitem[{{Tanaka} {et~al.}(2016){Tanaka}, {Wong}, {More}, {Dezuka}, {Egami}, {Oguri}, {Suyu}, {Sonnenfeld}, {Higuchi}, {Komiyama}, {Miyazaki}, {Onoue}, {Oyamada}, \& {Utsumi}}]{Tanaka2016}
{Tanaka}, M., {Wong}, K.~C., {More}, A., {et~al.} 2016, \apjl, 826, L19

\bibitem[{{Treu}(2010)}]{Treu2010}
{Treu}, T. 2010, \araa, 48, 87

\bibitem[{{Treu} {et~al.}(2022){Treu}, {Suyu}, \& {Marshall}}]{Treu2022}
{Treu}, T., {Suyu}, S.~H., \& {Marshall}, P.~J. 2022, \aapr, 30, 8

\bibitem[{{Tu} {et~al.}(2009){Tu}, {Gavazzi}, {Limousin}, {Cabanac}, {Marshall}, {Fort}, {Treu}, {P{\'e}llo}, {Jullo}, {Kneib}, \& {Sygnet}}]{Tu2009}
{Tu}, H., {Gavazzi}, R., {Limousin}, M., {et~al.} 2009, \aap, 501, 475

\bibitem[{{Ulmer} {et~al.}(2005){Ulmer}, {Adami}, {Covone}, {Durret}, {Lima Neto}, {Sabirli}, {Holden}, {Kron}, \& {Romer}}]{Ulmer2005}
{Ulmer}, M.~P., {Adami}, C., {Covone}, G., {et~al.} 2005, \apj, 624, 124

\bibitem[{{Vegetti} {et~al.}(2012){Vegetti}, {Lagattuta}, {McKean}, {Auger}, {Fassnacht}, \& {Koopmans}}]{Vegetti2012}
{Vegetti}, S., {Lagattuta}, D.~J., {McKean}, J.~P., {et~al.} 2012, \nat, 481, 341

\bibitem[{{Verde} {et~al.}(2019){Verde}, {Treu}, \& {Riess}}]{Verde2019}
{Verde}, L., {Treu}, T., \& {Riess}, A.~G. 2019, Nature Astronomy, 3, 891

\bibitem[{{Vernet} {et~al.}(2011){Vernet}, {Dekker}, {D'Odorico}, {Kaper}, {Kjaergaard}, {Hammer}, {Randich}, {Zerbi}, {Groot}, {Hjorth}, {Guinouard}, {Navarro}, {Adolfse}, {Albers}, {Amans}, {Andersen}, {Andersen}, {Binetruy}, {Bristow}, {Castillo}, {Chemla}, {Christensen}, {Conconi}, {Conzelmann}, {Dam}, {de Caprio}, {de Ugarte Postigo}, {Delabre}, {di Marcantonio}, {Downing}, {Elswijk}, {Finger}, {Fischer}, {Flores}, {Fran{\c{c}}ois}, {Goldoni}, {Guglielmi}, {Haigron}, {Hanenburg}, {Hendriks}, {Horrobin}, {Horville}, {Jessen}, {Kerber}, {Kern}, {Kiekebusch}, {Kleszcz}, {Klougart}, {Kragt}, {Larsen}, {Lizon}, {Lucuix}, {Mainieri}, {Manuputy}, {Martayan}, {Mason}, {Mazzoleni}, {Michaelsen}, {Modigliani}, {Moehler}, {M{\o}ller}, {Norup S{\o}rensen}, {N{\o}rregaard}, {P{\'e}roux}, {Patat}, {Pena}, {Pragt}, {Reinero}, {Rigal}, {Riva}, {Roelfsema}, {Royer}, {Sacco}, {Santin}, {Schoenmaker}, {Spano}, {Sweers}, {Ter Horst}, {Tintori}, {Tromp}, {van Dael}, {van der Vliet}, {Venema}, {Vidali}, {Vinther}, {Vola},
  {Winters}, {Wistisen}, {Wulterkens}, \& {Zacchei}}]{Vernet2011}
{Vernet}, J., {Dekker}, H., {D'Odorico}, S., {et~al.} 2011, \aap, 536, A105

\bibitem[{{Wang} {et~al.}(2022){Wang}, {Ca{\~n}ameras}, {Caminha}, {Suyu}, {Y{\i}ld{\i}r{\i}m}, {Chiriv{\`\i}}, {Christensen}, {Grillo}, \& {Schuldt}}]{Wang2022}
{Wang}, H., {Ca{\~n}ameras}, R., {Caminha}, G.~B., {et~al.} 2022, \aap, 668, A162

\bibitem[{{Weilbacher} {et~al.}(2020){Weilbacher}, {Palsa}, {Streicher}, {Bacon}, {Urrutia}, {Wisotzki}, {Conseil}, {Husemann}, {Jarno}, {Kelz}, {P{\'e}contal-Rousset}, {Richard}, {Roth}, {Selman}, \& {Vernet}}]{Weilbacher2020}
{Weilbacher}, P.~M., {Palsa}, R., {Streicher}, O., {et~al.} 2020, \aap, 641, A28

\bibitem[{{Wong} {et~al.}(2020){Wong}, {Suyu}, {Chen}, {Rusu}, {Millon}, {Sluse}, {Bonvin}, {Fassnacht}, {Taubenberger}, {Auger}, {Birrer}, {Chan}, {Courbin}, {Hilbert}, {Tihhonova}, {Treu}, {Agnello}, {Ding}, {Jee}, {Komatsu}, {Shajib}, {Sonnenfeld}, {Blandford}, {Koopmans}, {Marshall}, \& {Meylan}}]{Wong2020}
{Wong}, K.~C., {Suyu}, S.~H., {Chen}, G. C.~F., {et~al.} 2020, \mnras, 498, 1420

\end{thebibliography}
\end{document}